\documentclass[12pt]{iopart} 
\usepackage{iopams} 
\usepackage{amssymb}
\usepackage{graphicx} 
\usepackage{amsfonts}
\usepackage{appendix}
\usepackage{float}
\usepackage{subfig}
\newcommand{\BEA}{\begin{eqnarray}}
\newcommand{\EEA}{\end{eqnarray}}
\newcommand{\nn}{\nonumber \\}
\newcommand{\x}{\hat x}
\newcommand{\p}{\hat p}
\newcommand{\ha}{\hat a}
\newcommand{\had}{\hat a^\dagger}
\newcommand{\sq}{\sqrt{2}}
\newcommand{\half}{\frac{1}{2}}
\newcommand{\D}{\hat D}
\newcommand{\hsr}{\hat S \left(s\right)}
\newcommand{\hsrm}{\hat S \left(- s\right)}
\newcommand{\om}{\omega}
\newcommand{\Om}{\Omega}

\newcommand{\hS}{\hat S}
\newcommand{\z}{\zeta}

\newcommand{\te}{\theta}
\newcommand{\hH}{\hat H}
\newcommand{\hV}{\hat V}
\newcommand{\hroi}{\hat \rho_{\textnormal I}}
\newcommand{\bmui}{\bi{\bmu}_{\textnormal I}}
\newcommand{\vi}{\hat V_{\textnormal I}(t)}

\newcommand{\ui}{\hat U_{\textnormal I}(t)}
\newcommand{\hA}{\hat A}

\newcommand{\hU}{\hat U_0}
\newcommand{\vac}{\left \vert 0 \right \rangle}
\newcommand{\lac}{\left\langle 0 \right\vert}
\newcommand{\hro}{\hat{\rho}}
\newcommand{\schro}{Schr\"{o}dinger}
\newcommand{\heis}{\textnormal{Heisenberg}}
\newcommand{\inte}{\textnormal{Interaction}}
\newcommand{\m}{\frac{\mu_x}{\sqrt{2}}}
\newcommand{\gO}{\frac{g}{\Omega}}
\newcommand{\Wsp}{W^{\textnormal{SP}}}
\newcommand{\Whp}{W^{\textnormal{HP}}}
\newcommand{\Wip}{W^{\textnormal{IP}}}
\newcommand{\K}{\hat{K}}
\begin{document}

\title[Visualizing the quantum interaction picture in phase space]{Visualizing the quantum interaction picture in phase space}
\author{Bahar Mehmani$^1$, Andrea Aiello$^{1,2}$}
\address{$^1$ Max Planck Institute for the Science of Light, G\"{u}nter-Scharowsky-Stra$\ss$e 1/Bau 24, D-91058 Erlangen, Germany}
\address{$^2$ Institute for Optics, Information and Photonics, University Erlangen-N\"{u}rnberg, Staudtstra$\ss$e 7/B2, D-91058 Erlangen, Germany}
\ead{bahar.mehmani@mpg.mpl.de} 

\begin{abstract}
We illustrate the correspondence between the quantum \inte\ Picture-evolution of the state of a quantum system in Hilbert space and a combination of local and global transformations of its Wigner function in phase space. To this aim, we consider the time-evolution of a quantized harmonic oscillator driven by both a linear and a quadratic (in terms of bosonic creation and annihilation operators) potentials and employ the Magnus series to derive the exact form of the time-evolution operator. In this case, the Interaction Picture corresponds to a local transformation of phase space-reference frame into the one that is co-moving with the Wigner function. 
\end{abstract}
\submitto{\NJP}
\maketitle
 \section{Introduction}\label{intro}
One of the most important lessons that one learns in quantum mechanics courses is to choose a proper picture  for describing the dynamics of a system. An appropriate picture is the one in which the physical properties of the system can be easily revealed and the mathematical calculations involved are relatively simple. In general, there are three pictures for this aim: the \schro, the \heis, and the \inte\ picture~\cite{Parisi}. The \schro\  picture (SP) is more suitable for studying closed and conservative systems. In this picture the only time-dependent quantity of the system is its quantum state. Correspondingly, observables of the system are constant in time. However, the number of situations in which a quantum system can be considered as a closed one is limited and, furthermore, most of the interesting phenomenon in the quantum world such as decoherence and optical cooling, to just name two, happen when a quantum system couples to its surrounding environment, to an external force or to both of them. Hence, the Hamiltonian describing the system is more complicated due to the presence of the environment and/or the external force. In many of such situations it is impossible to find an exact solution to the \schro\ equation. However, sometimes, adopting the \heis\ picture (HP) makes it possible to find the time-evolution of the expectation values of the observables of the system. In HP the state does not vary with time while the observables are time-dependent and their evolution is described by the \heis\ equation of motion. 
Finally, the last scheme, namely the \inte\ picture (IP) one, is the most suitable picture when the Hamiltonian of a system can be written as a sum of two parts: a time-independent term of which the eigenstates and eigenenergies are known, and a usually time-dependent term which influences the dynamics of the system. Employing this picture enables one to set the dynamics arising from the time-independent Hamiltonian aside and focus on the influence of its time-dependent part on the evolution of the system. That is why this picture is frequently employed in quantum optics where  matter interacts with the radiation field. 

All these three pictures are described in terms of unitary transformations in Hilbert space. The aim of this work is to map such transformations in phase space and illustrate the simplicity of the IP time-evolution of a quantum system in comparison to its SP counterpart. Although SP and HP have been mapped into classical active and passive transformations~\cite{Parisi}, to the best of our knowledge, there is no such direct comparison for the interaction picture in the physics literature.
Here we fill this gap and show that by taking active local and global transformations one can transform the IP to the SP or HP. In this way, it is easier to understand the concept of performing different unitary transformations in Hilbert space and its consequences on the evolution of the state of the system of interest.

To this end, we take a single harmonic oscillator as our quantum system, initially prepared in an ideal squeezed state~\cite{Mandel} and first study its free time-evolution in SP and HP. Then by employing the Wigner-Weyl description we calculate the time-evolution of its corresponding Wigner function. The basic advantage of the Wigner-Weyl representation~\cite{Case08,Hillery,Agarwal,Schleich} is that the operators $\x$ and $\p$ turn into $c$-numbers  making the correspondence with canonical transformations possible. The Wigner function of the coherent and squeezed states of a quantized harmonic oscillator is a Gaussian function of these $c-$number variables and its corresponding phase space distribution can be pictured by circles and ellipses, respectively~\cite{Kim90,Kim88,Kim89}. This is the subject of sections~\ref{Sec1:HO} and~\ref{Sec:Passive}.
In section~\ref{Sec:Interaction} we introduce two types of a time-dependent potential to the system: a linear and a quadratic term (in terms of the bosonic creation and annihilation operators). Then we take the IP to study the dynamics of the initially squeezed state under the driving potentials.  We explicitly show that the IP- time-evolution of the state in Hilbert space corresponds to a transformation to a local frame that is co-rotating with the Wigner function in phase space. Furthermore, we show switching from one picture to another is equivalent to performing different active and/or passive transformations in phase space.

\section{Basics of the quantized harmonic oscillator}\label{Sec1:HO}

A single mode harmonic oscillator with unit mass and frequency $\om_0$ is described by the sum of its kinetic and potential energy
\BEA\label{Hconv}
\hH_0 = \frac{ \om_0}{2} (\hat p^2 + \hat x^2).
\EEA
Here $\hat x$ and $\hat p$ represent quadratures of the quantized harmonic oscillator and are defined in terms of the annihilation operator $\ha$ and  the creation operator $\had$ as
\numparts
\BEA\label{quad-def}
\hat x & = \frac{1}{\sqrt{2}} ( \hat a + \hat a^\dagger ),\\
\hat p & = \frac{1}{\rmi \sqrt{2}}  ( \hat a - \hat a^\dagger ),
\EEA
\endnumparts
where $\hbar$ is set to one. 
Alternatively, this system may be described in terms of $\ha$ and $\had$ as
\BEA\label{HDirac}
\hat H_0 =\om_0 \left(\hat a^\dagger \hat a + \frac{1}{2}\right),
\EEA
where the commutator of the creation and the annihilation operator is $[\ha, \had]= 1$. 
$\hH_0$ has the energy eigenvalues $E_n$ defined as
\BEA\label{En}
E_n =\om_0 \left(n+ \frac{1}{2}\right), \textnormal{with} \qquad n = 0, 1,2, \cdots.
\EEA
The energy eigenstates (number states) $\vert n \rangle$ such that:
\BEA\label{n}
\hH_0 \vert n \rangle = E_n \vert n \rangle.
\EEA
Given $\hH_0$, the time-evolution of quadrature operators $\x$ and $\p$ is given by the Heisenberg equation of motion as
\numparts
\BEA\label{x,p:t}
\x(t) & =\x(0) \cos (\om_0 t)  + \p(0) \sin (\om_0 t) ,\\
\p(t) & = - \x(0)  \sin (\om_0 t) + \p(0) \cos (\om_0 t) .\label{p,x:t}
\EEA
\endnumparts

Any superposition of the number states is a solution to the Schr\"{o}dinger equation for a quantized harmonic oscillator described by the Hamiltonian $\hH_0$. There are two classes of such superpositions that are of great interest: coherent and squeezed states.
A coherent state $\vert \alpha \rangle$ is described as the displaced vacuum state $\vert 0 \rangle$ and is generated by a canonical transformation of the vacuum state $\vac$ in phase space
\BEA\label{cohstatevac}
\vert \alpha \rangle = \D(\alpha) \vert 0 \rangle,
\EEA
where $\D(\alpha)$ is the unitary displacement operator defined as~\cite{Glauber,Scully}
\BEA
\label{D_def}
\D(\alpha)= e^{( \alpha \hat a^{\dagger} - \alpha^* \ha)}.
\EEA
The displacement parameter $\alpha$, is a complex number and may be decomposed to
\BEA\label{alpha-def}
\alpha \equiv \frac{1}{\sq} ( a + \rmi b),
\EEA

Squeezed states are of interest because they provide reduced fluctuations in one quadrature, when compared with the coherent state. A squeezed state is described by the unitary squeezing operator~\cite{Mandel,Yuen76,Caves81,Walls83,Tapia93} $\hS(\z)$ in a similar fashion. However, the difference lies in the fact that $\hS(\z)$ is not linear in $\ha$ and $\had$ and is defined as 
\BEA\label{SqueeOp-def}
\hS(\z) = \exp \left[ \left(  \z^* \ha^2- \z  {\ha}^{\dagger 2}\right)/2 \right],\qquad \z = s e^{\rmi \te},
\EEA
where $s = \vert \z \vert$ is the strength of the squeezing and $\te = \arg(\z)$ determines the direction along which the squeezing is performed. In phase space the squeezed states correspond to Gaussian distributions with unequal widths as opposed to the symmetric distribution of the coherent state.~\cite{Kim88}

Throughout this paper, for the sake of simplicity and without the loss of generality, we take both the displacement parameter $\alpha$ and the squeezing parameter $\z$ as real numbers.  Moreover, we assume the squeezed state to be an ideal squeezed state~\cite{Mandel}. 

Since we want to find a classical correspondence for the unitary transformations in the Interaction picture, we need to map the time-evolution of the ideal squeezed state mentioned above into classical phase space and study its time-evolution. This is best described with the help of the Wigner function and Weyl transform~\cite{Case08,Hillery,Agarwal}.
Knowing $\hro(t)$ the state of the quantum system at any time, we can build up the symmetric characteristic function~\cite{Barnett} $\Tr[\hro(t) \D(\chi)]$ and from there the corresponding SP-Wigner function $W(x,p;t)$ as
\BEA\label{Wignerdef}
W(x,p;t) = \frac{1}{(2 \pi)^2} \int \rmd\xi \int \rmd\eta\,\rme^{-\rmi (\xi x + \eta p)}\,\Tr[\hro(t) \D(\chi)],
\EEA
where $\D(\chi)$ is the displacement operator with the parameter $\chi$ defined as
\BEA\label{chi-def}
\chi \equiv \frac{1}{\sq} (- \eta +\rmi \xi).
\EEA
Both integrations are from $-\infty$ to $\infty$. 

\section{Schr\"{o}dinger and Heisenberg pictures vs. passive and active transformations}\label{Sec:Passive}

Before embarking on the classical correspondence of the IP, let us first explicitly show for a single-mode harmonic oscillator how SP and HP map to active and passive transformations in phase space, respectively.
For a harmonic oscillator with the Hamiltonian given by (\ref{HDirac}), the unitary time-evolution operator $\hU$, which governs the evolution of the state, reads
\BEA\label{U0}
\hU(t) = \rme^{- \rmi \hH_0 t} = \rme^{- \rmi \om_0 t (\had \ha + \half)}.
\EEA
If the system is initially in an ideal squeezed state, 
\BEA\label{int_ro_ini}
\hro(0) = \D(\m) \hsr \vac \lac \hsrm \D(- \m),
\EEA
with $\mu_x$ representing the displacement along the $x$-axis in phase space and $s$ denoting the real squeezing parameter, then the time-evolution of the state is given by
\begin{equation}
\label{rhot_sq-free}
\hro(t) = \hat U_0(t) \hro(0) \hat U^\dagger_0(t).
\end{equation}
\begin{figure}
\subfloat[$\om_0t=0$]{\label{fig:12}\includegraphics[width=0.3\textwidth]{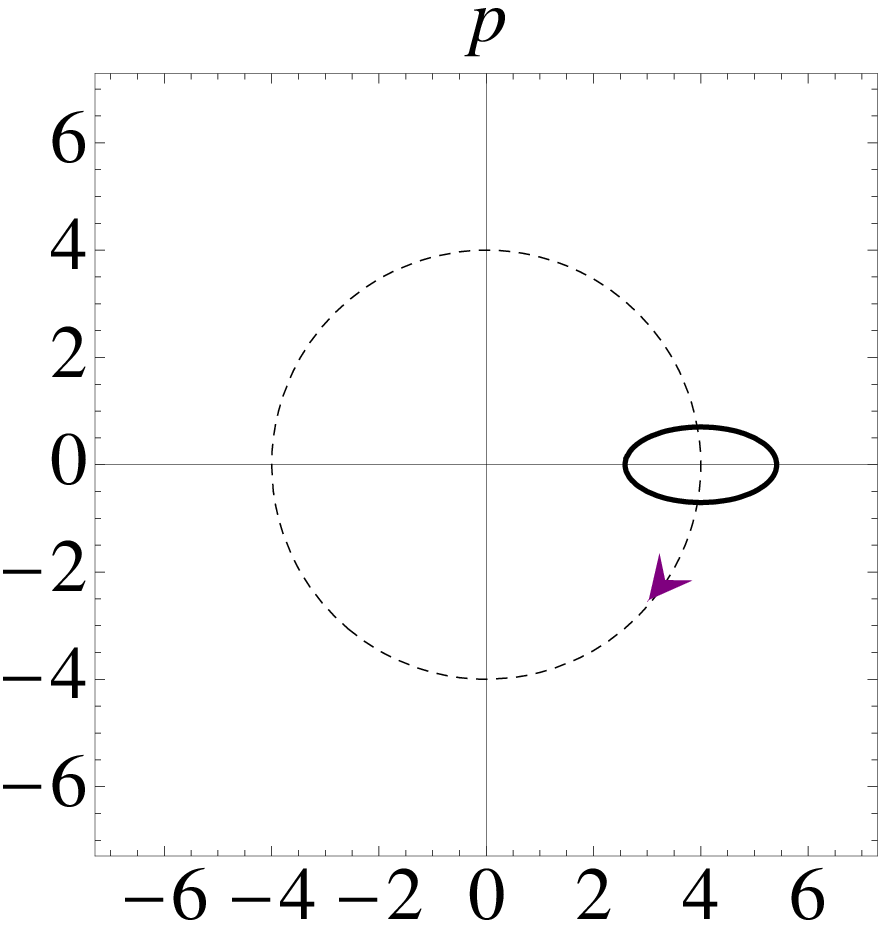}}
\subfloat[$\om_0t=\frac{\pi}{4}$]{\label{fig:13}\includegraphics[width=0.3\textwidth]{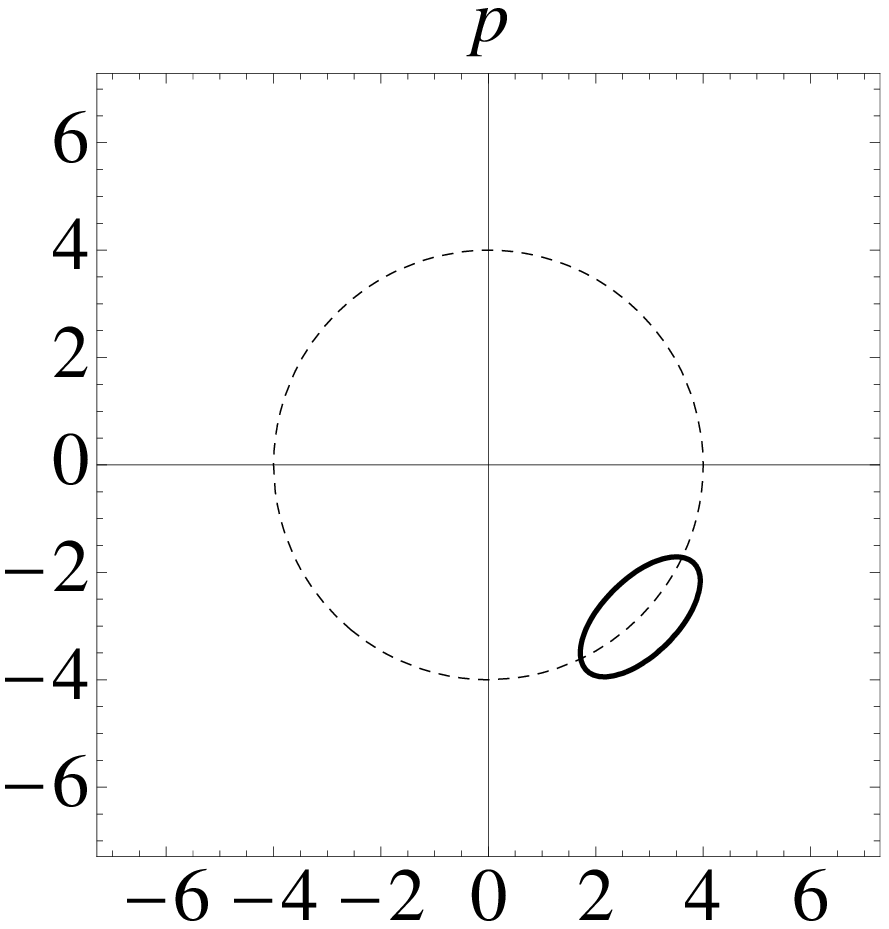}}
\subfloat[$\om_0t=\frac{\pi}{2}$]{\label{fig:14}\includegraphics[width=0.313\textwidth]{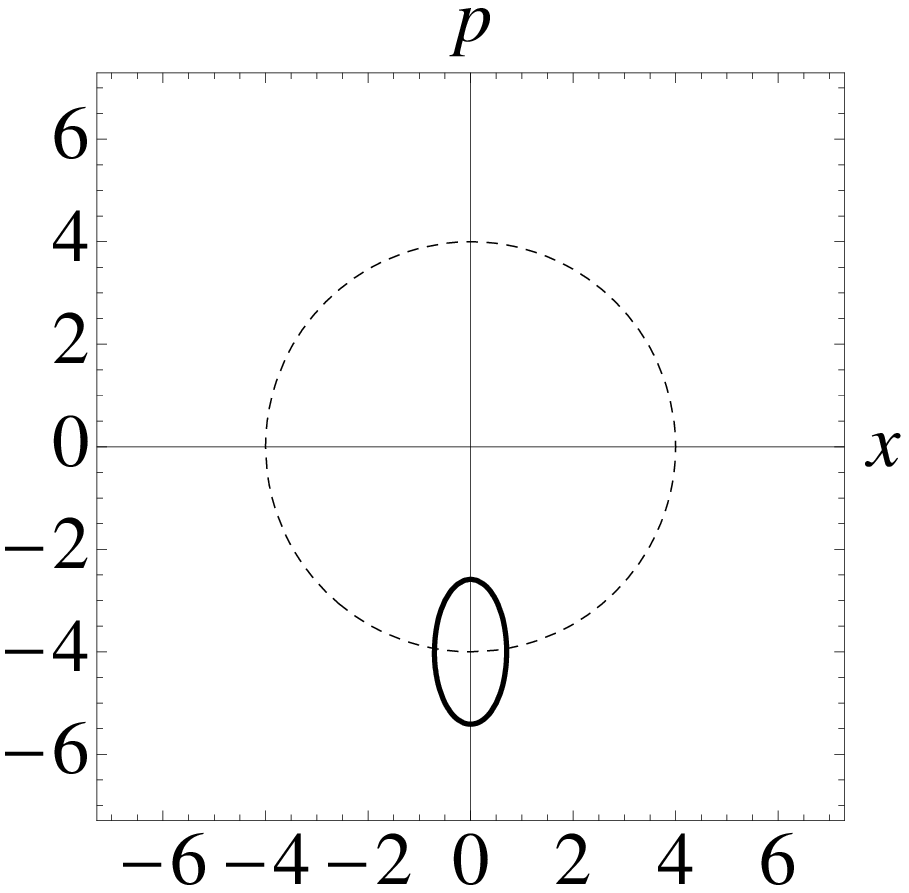}}\\
\subfloat[$\om_0t=\frac{3 \pi}{4}$]{\label{fig:15}\includegraphics[width=0.3\textwidth]{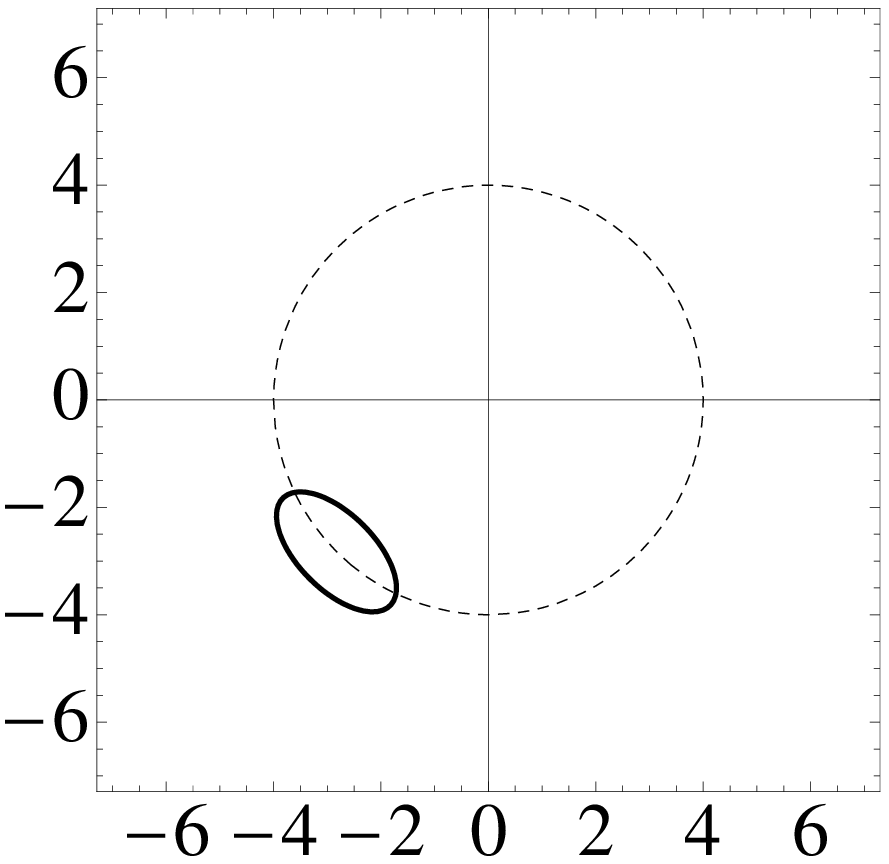}}
\subfloat[$\om_0t=\pi$]{\label{fig:16}\includegraphics[width=0.3\textwidth]{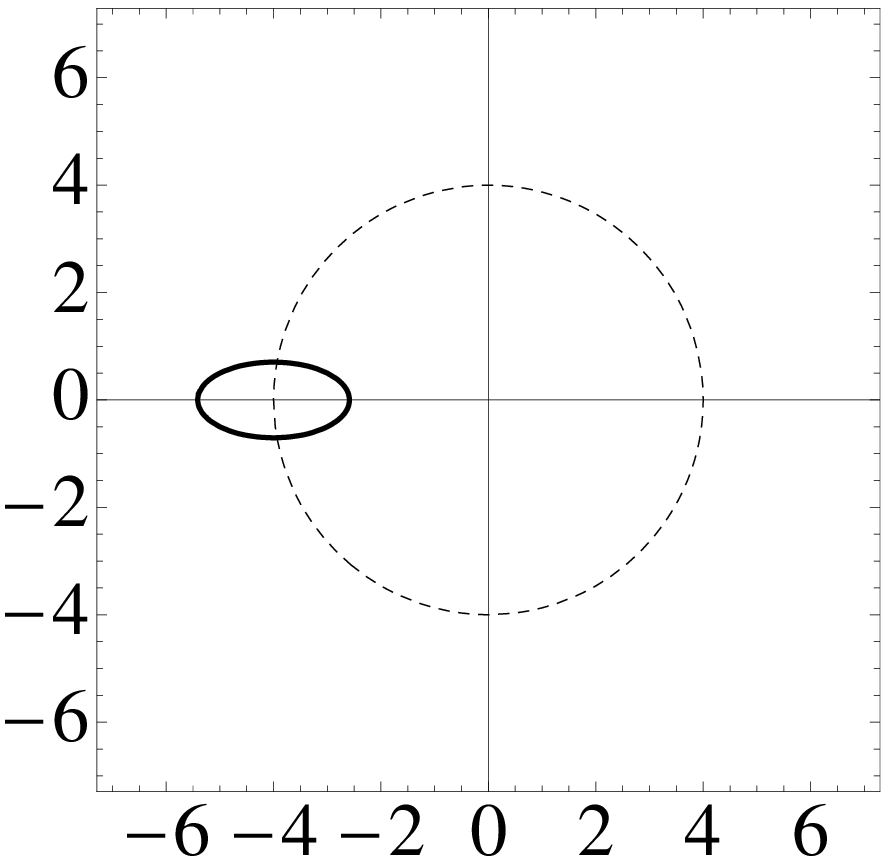}}
\subfloat[$\om_0t=\frac{5\pi}{4}$]{\label{fig:17}\includegraphics[width=0.313\textwidth]{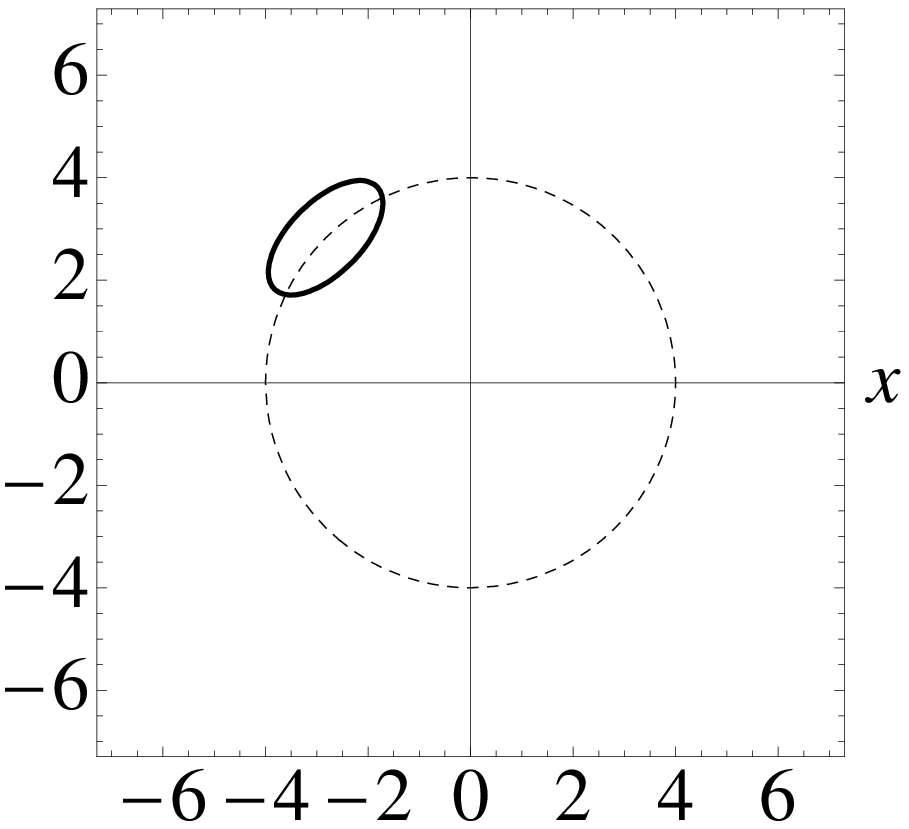}}\\
\subfloat[$\om_0t=\frac{3 \pi}{2}$]{\label{fig:18}\includegraphics[width=0.3\textwidth]{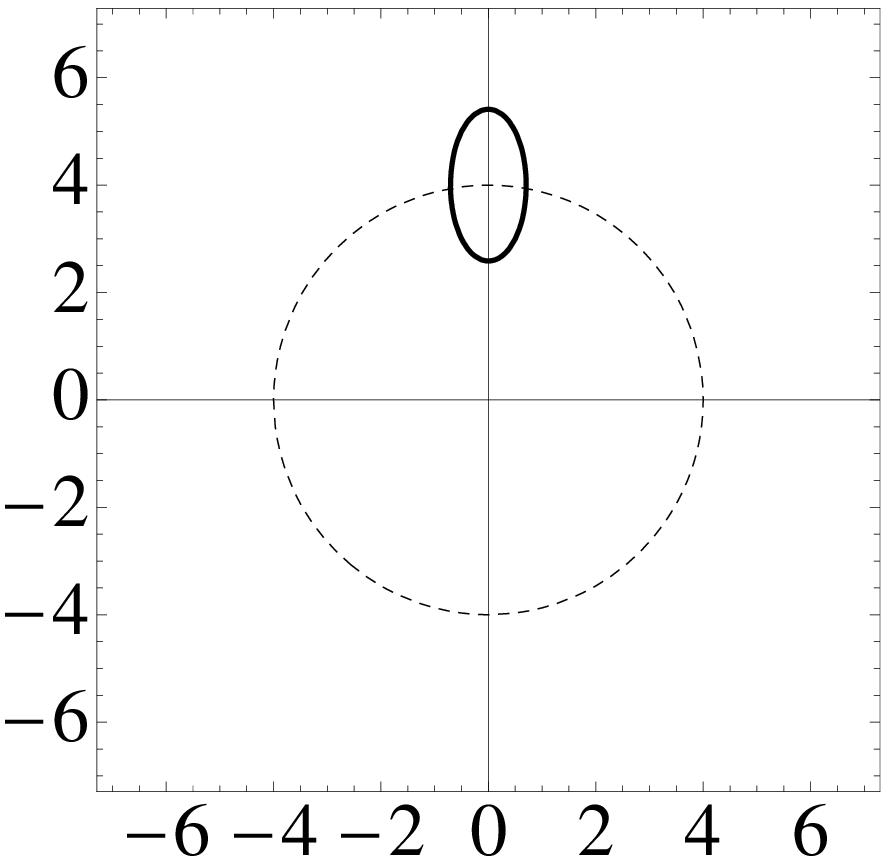}}
\subfloat[$\om_0t=\frac{7\pi}{4}$]{\label{fig:19}\includegraphics[width=0.3\textwidth]{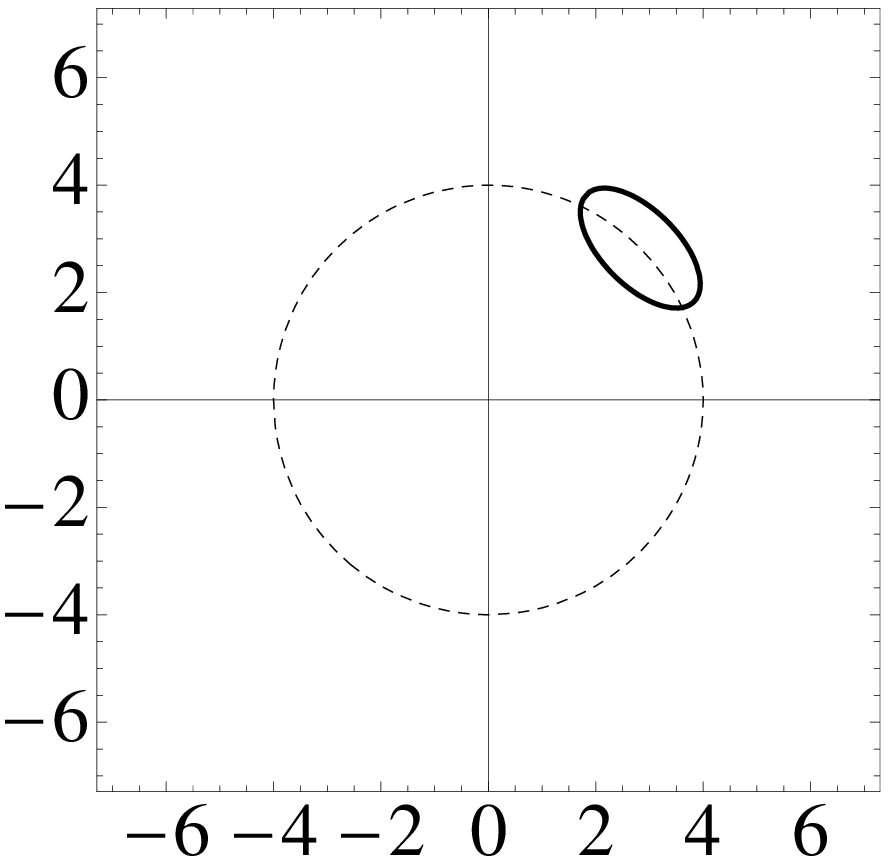}}
\subfloat[$\om_0t=2 \pi$]{\label{fig:20}\includegraphics[width=0.313\textwidth]{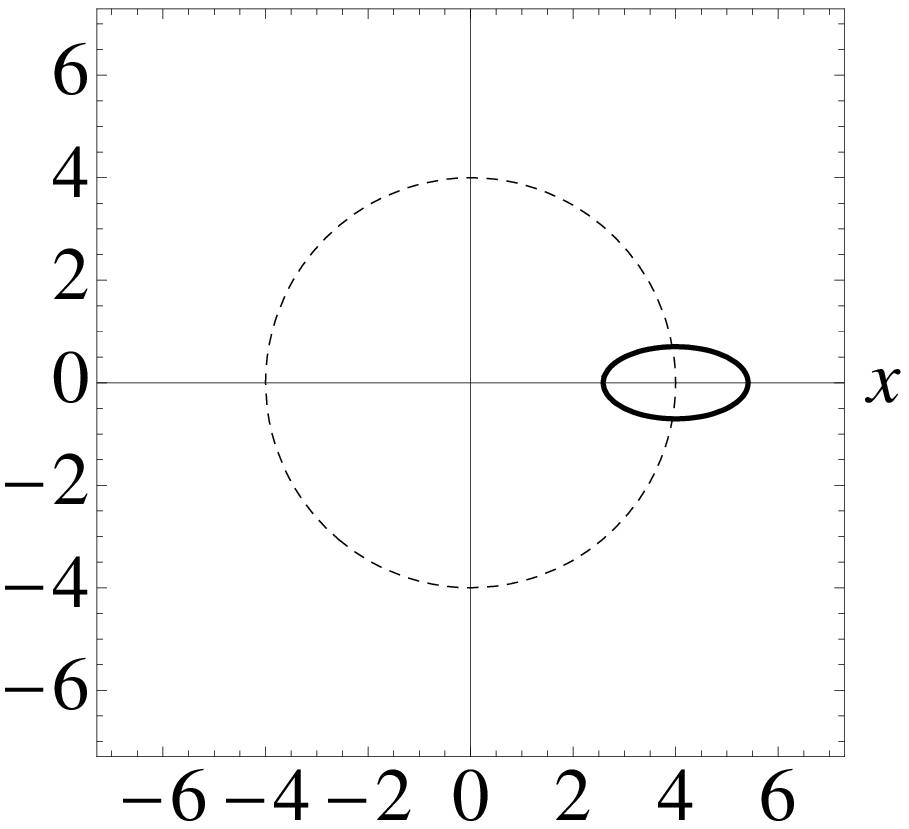}}
\caption{Free time-evolution of the $\frac{1}{\rme}$-contour of the SP-Wigner function of an ideal squeezed state given by (\ref{Wshro_final}) in units of $\om_0 t$ ($\sigma_x = 1, \sigma_p = \half,\mu_x = 4$, in arbitrary units). The evolution is a combination of a local counterclockwise rotation of the ellipse about its centroid through angle $\om_0 t$, and the global clockwise rotation of its center about the origin of the phase space on the dashed-line circle through the same angle.}
\label{fig.1}
\end{figure}
Knowing $\hro(t)$, the Wigner function at time $t$ can be derived by first calculating the symmetric characteristic function as the expectation value of the displacement operator $\D(\chi)$ as
\BEA\label{trfin_sq-free}
\fl \Tr\left[\hro(t) \D(\chi)\right]&= \exp{\left[\rmi \xi  \mu_x \cos (\om_0 t) - \rmi \eta \mu_x \sin (\om_0 t) \right]}
\exp{\left[- \frac{\xi^2}{4}  \left( \frac{\cos^2(\om_0 t)}{2 \sigma^2_p} + \frac{\sin^2 (\om_0 t)}{2 \sigma^2_x}\right)\right]}\nn
 & \times\exp{\left[- \frac{\eta^2}{4}  \left(\frac{\cos^2(\om_0 t)}{2 \sigma^2_x} + \frac{\sin^2 (\om_0 t)}{2 \sigma^2_p} \right)\right]} \exp{\left[ - \frac{\xi \eta}{8} \sin (2 \om_0 t) \left( \frac{1}{\sigma^2_x}- \frac{1}{\sigma^2_p}  \right)\right] }.\nn
\EEA
where we substitute $e^{- 2 s} = 2\sigma^2_x$ and $e^{ 2 s} =2 \sigma^2_p$ in the coefficients of $\xi^2$,  $\eta^2$ and $\xi \eta$. 
Substituting (\ref{trfin_sq-free}) into the definition of the Wigner function given by (\ref{Wignerdef}) and performing two Gaussian integrals over $\xi$ and $\eta$ then yields to the SP-Wigner function as
\BEA
\label{Wshro_final}
\fl \Wsp_0(x,p;t) = \frac{1}{\pi} \exp{\left\{ - \half \left[\bi r - R(-\om_0\,t) \bmu \right]^T \cdot \Gamma^{- 1}(t) \cdot \left[\bi r - R(-\om_0\,t) \bmu \right]\right\}},
\EEA
where the superscripts ``$T$'' and ``${-1}$'' denote matrix transposition and inversion operations, respectively. The suffix $\textnormal{SP}$ denotes the Schr\"{o}dinger picture and the index $0$ in $W_0(x,p;t)$ represents the fact that the system evolves freely in time.
The vectors $\bmu$ and $\bi r$ represents the initial displacement of the Wigner function and the coordinates of phase space, respectively, as
\BEA
\bmu  = \left(\begin{array}{c}\mu_{x} \\0\end{array}\right),\label{mu_def} \qquad
\bi r= \left(\begin{array}{c}x \\p\end{array}\right).\label{vec_r_def}
\EEA
The matrix $\Gamma^{- 1}(t)$ is defined as the clockwise rotation of the squeezing matrix through the angle $\om_0 t$
\BEA\label{Gamma_def}
\Gamma^{-1}(t) \equiv R(- \om_0 t)\cdot \Xi^{-1} \cdot R( \om_0 t),
\EEA
where the diagonal $2\times2$ squeezing matrix $\Xi^{-1}$ is given by 
\BEA\label{covmatrix_int}
\Xi^{-1}\equiv \left(\begin{array}{cc}\frac{1}{\sigma^2_x} & 0 \\0 &\frac{1}{\sigma^2_p}\end{array}\right).
\EEA
Notice that the matrix $R(\om_0 t)$ defines a counterclockwise rotation by $\om_0 t$ as
\BEA\label{Rotation_def}
R(\om_0 t) = \left(\begin{array}{cc}\cos\om_0 t &  - \sin\om_0 t \\ \sin\om_0 t & \cos\om_0 t\end{array}\right).
\EEA 

The free-time evolution of the $\frac{1}{\rme}$-contour of the SP-Wigner function of an ideal squeezed state is represented in figure~\ref{fig.1}. As it is illustrated in figures~\ref{fig:12}-\ref{fig:20}, the elliptical distribution performs two types of rotations: ${\it i)}$ a global clockwise rotation (active transformation) of the distribution about the origin of the phase space on a circular path defined by the initial amount of displacement; ${\it ii)}$ a local counterclockwise rotation  of the the squeezing direction through the same angle $\om_0 t$ about its centroid. From the mathematical point of view, the global rotation is produced by the clockwise rotation matrix $R(-\om_0t)$ acting on the vector $\bmu$, whereas the action of the  rotation matrix $R(\om_0 t)$ on the squeezing matrix $\Xi^{-1}$ amounts to its local counterclockwise rotations.

In (\ref{Wshro_final}), if we take $R(- \om_0 t)$ from left in to the first bracket and $R(\om_0 t)$ from right in to the second bracket, the resulting Wigner function can be written in terms of a new phase space coordinate which is described by
\begin{equation}\label{r-r'}
\bi{r'} = R(\om_0 t)\, \bi{r}.
\end{equation}
In this new local frame, attached to the centroid of the Wigner function, the local counterclockwise rotation of the squeezing direction of the Wigner function about its centroid cancels out its global clockwise rotations about the origin
\BEA\label{W':HP}
\Whp_0(x',p') = \frac{1}{\pi} \exp{\left[- \half \left(\bi{r'} -  \bmu \right)^T\, \cdot \Xi^{-1}\cdot \, \left(\bi{r} - \bmu \right)\right]}.
\EEA
The passive transformation (\ref{r-r'}) corresponds to changing the reference frame to the one in which $x(t)$ and $p(t)$ are rotated back to their initial values $x(0)$ and $p(0)$. The suffix \mbox{HP} represents that the Wigner function is described in the Heisenberg picture, in which the Wigner function is  fixed in time while the operators $\x$ and $\p$ evolve according to (\ref{x,p:t}) and (\ref{p,x:t}). 
Hence we see that the passive transformation (\ref{r-r'}) in phase space corresponds to the Heisenberg picture.

The main message of this section is that the action of the unitary transformation $\hU$ on an ideal squeezed state in the Hilbert space corresponds to two types of rotations of its Wigner function in the phase space: ${\it i)}$ a global clockwise rotation about the origin through angle $\om_0 t$, ${\it ii)}$ a local counterclockwise rotation of the Wigner function through angle $\om_0 t$ about its centroid. Further, the action of $\hU$ on the system in Hilbert space is equivalent to a counterclockwise rotation of the coordinate system with angle $\om_0 t$, or equivalently, performing a passive transformation in phase space.

\section{Interaction Picture and mixed rotations}\label{Sec:Interaction}

The interaction picture is the most frequently employed picture in quantum dynamics. Specifically, it is the preferred picture in quantum optics where a system is driven by the time-dependent electric field of light. 

Considering the single mode harmonic oscillator described in section~\ref{Sec1:HO}, we now assume it undergoes a time-dependent potential $\hV(t)$. Thus the total Hamiltonian is given by 
\begin{equation}\label{interaction}
\hH = \hH_0 + \hV(t),
\end{equation}
with $\hH_0$ being defined in (\ref{HDirac}). $\hV(t)$ influences the temporal behavior of the system and is a linear driving potential described as
\begin{equation}\label{H1_def}
\hV(t) =  g (\rme^{- \rmi \om_1 t} \alpha \ha + \rme^{\rmi \om_1 t} \alpha^* \had),
\end{equation}
where $g$ is a real coupling constant. Physically, this corresponds to a forced harmonic oscillator with the driving frequency $\om_1$.

In the IP both the state and observables are time-dependent. The observables evolve in time according to the Heisenberg equation of motion as
\begin{equation}
\label{Sch-Int-op}
\hat {\Or}_{\textnormal I}(t) = \hat U^\dagger_0(t) \, \hat{\Or} \, \hat U_0(t),
\end{equation}
where $\hat U_0(t)$ is given by (\ref{U0}). 
Accordingly, the time-dependent potential in the interaction picture, $\hV_{\textnormal I}$, reads as
\begin{equation}\label{VI}
\hV_{\textnormal I}(t) = g (e^{- \rmi \Om t} \alpha \ha + e^{\rmi \Om t} \alpha^* \had),
\end{equation}
where $\Omega$ is defined as the sum of the natural frequency $\om_0$ and the driving frequency $\om_1$:
\begin{equation}\label{Omega}
\Om \equiv \om_0 + \om_1.
\end{equation}
When $\Om = 0\ (\om_1 = - \om_0)$, the interaction picture potential becomes time-independent. We shall study this case in more details in section~\ref{sec:Omzero}. For the moment we assume $\Om \neq 0$.

The time evolution of the state is described by the interaction picture unitary transformation $\ui$ as
\begin{equation}\label{int-w-sol}
\hro_{\textnormal I}(t) = \ui \hro(0) \hat U^\dagger_{\textnormal I}(t),
\end{equation}
where $\ui$ is given by the Magnus series~\cite{Magnus}
\begin{equation}\label{UMagnus}
\ui = \exp{\left\{ - \rmi \left[\hat A_1(t) + \hat A_2(t) + \hat A_3(t) + \cdots \right]\right\}}.
\end{equation}
The first three terms in the exponential are given by
\numparts
\BEA\label{Magnus}
\fl\hat A_1(t) =  \int_0^t \rmd t_1\, \hat V_{\textnormal I}(t_1),\label{A1Mag}\\
\fl\hat A_2(t) = \frac{1}{2! \rmi } \int_0^t \rmd t_2 \int_0^{t_2} \rmd t_1\, \left[ \hat V_{\textnormal I}(t_1), \hat V_{\textnormal I}(t_2)\right],\\\label{A2Mag}
\fl\hat A_3(t) = \frac{1}{3! (\rmi^2 )} \int_0^t \rmd t_3 \int_0^{t_3} \rmd t_2 \int_0^{t_2} \left\{ \left[\hat V_{\textnormal I}(t_1),\left[ \hat V_{\textnormal I}(t_2), \hat V_{\textnormal I}(t_3\right] \right] + \left[\hat V_{\textnormal I}(t_3), \left[ \hat V_{\textnormal I}(t_2), \hat V_{\textnormal I}(t_1)\right] \right]\right\}\, \rmd t_1.\nn
\label{A3Mag}
\EEA
\endnumparts
where $\vi$ is given by (\ref{VI}).
Thus the time evolution of the state of the system can be derived by calculating $\hat A_1(t), \hat A_2(t), \dots$, and subsequently constructing
 $\ui$, which ultimately enables us to calculate $\rho_{\textnormal I}(t)$. Then it is straight forward to calculate the IP-Wigner function and present a classical correspondence to $\ui$. We will ultimately transform the IP-density matrix back to the \schro\ picture $\hro_{\textnormal{S}}(t)$ with the help of $\hat U(t)$ as
\begin{equation}\label{Int-Sch-state}
\hro_{\textnormal{S}}(t) = \hat U_0(t)  \hro_{\textnormal I}(t) \hat U^\dagger_0(t),
\end{equation}
and study the corresponding SP-Wigner function.

For the first two term $\hA_1(t)$, and $\hA_2(t)$ we have
\BEA\label{A1}
\hat A_1(t) = \rmi \left[ \nu(t) \had - \nu^*(t) \ha +\gO\left(\alpha^* \had - \alpha \ha \right)\right],
\EEA
where $\nu(t)$ is defined as
\begin{equation}\label{nu}
\nu(t) \equiv - \frac{g}{\Om} \alpha^* e^{\rmi \Om t},
\end{equation}
with $\alpha^{*}$ being the complex conjugate of the displacement parameter defined in (\ref{alpha-def}).
Since the commutator $[\ha,\had] = 1$, the commutator of $\vi$ at two different times turns into a purely imaginary function of time. Hence the higher order terms in Magnus series given by (\ref{Magnus}) vanish and we are able to write down the exact solution for the equations of motion. This conclusion justifies our choice of $\hH_1(t)$ and enables us to illustrate the classical correspondence of the IP in phase space.
For $A_2(t)$ we get
\BEA\label{A2}
A_2(t) = \frac{g^2 \vert \alpha \vert^2}{\Om^2} \left(\Om t - \sin\Om t \right).
\EEA
As a result $\ui$ is given by
\BEA\label{UI}
\ui = \rme^{- \rmi \frac{g^2\vert \alpha \vert^2}{\Om^2}\left( \Om t - 2 \sin\Om t \right)} \D \left(\nu(t)\right) \D (\gO \alpha^*).
\EEA

\begin{figure}
\subfloat[$\Om t = 0$]{\label{fig:22}\includegraphics[width=0.3\textwidth]{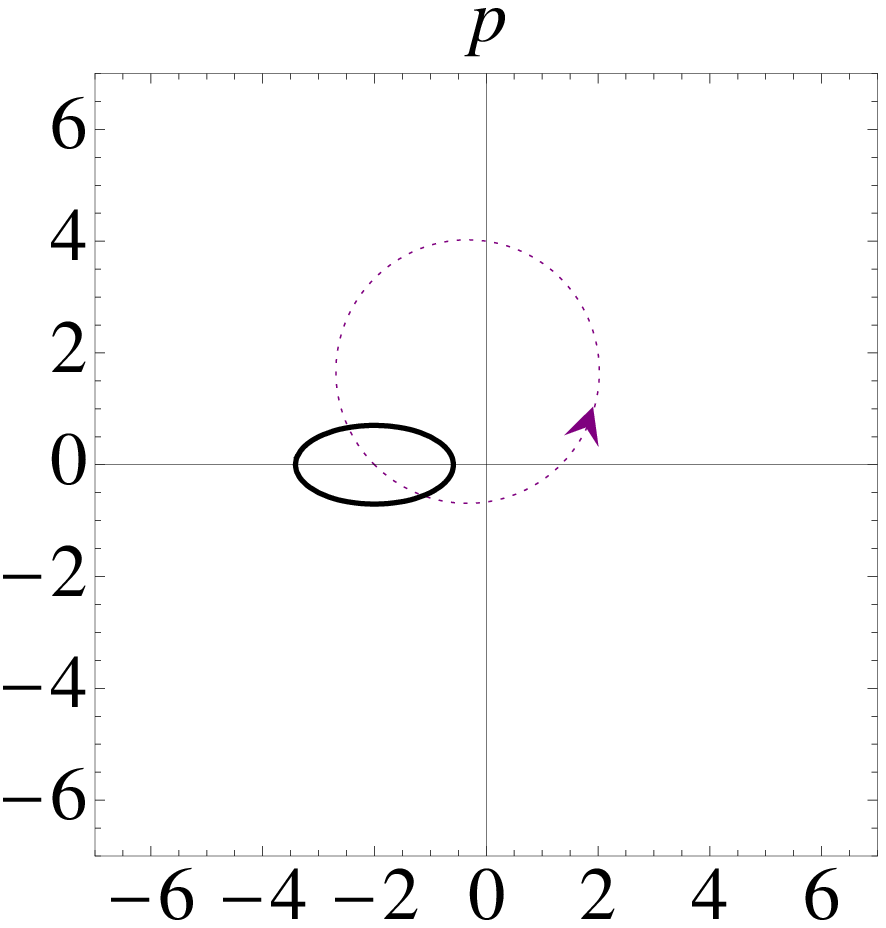}}
\subfloat[$\Om t=\frac{\pi}{4}$]{\label{fig:23}\includegraphics[width=0.3\textwidth]{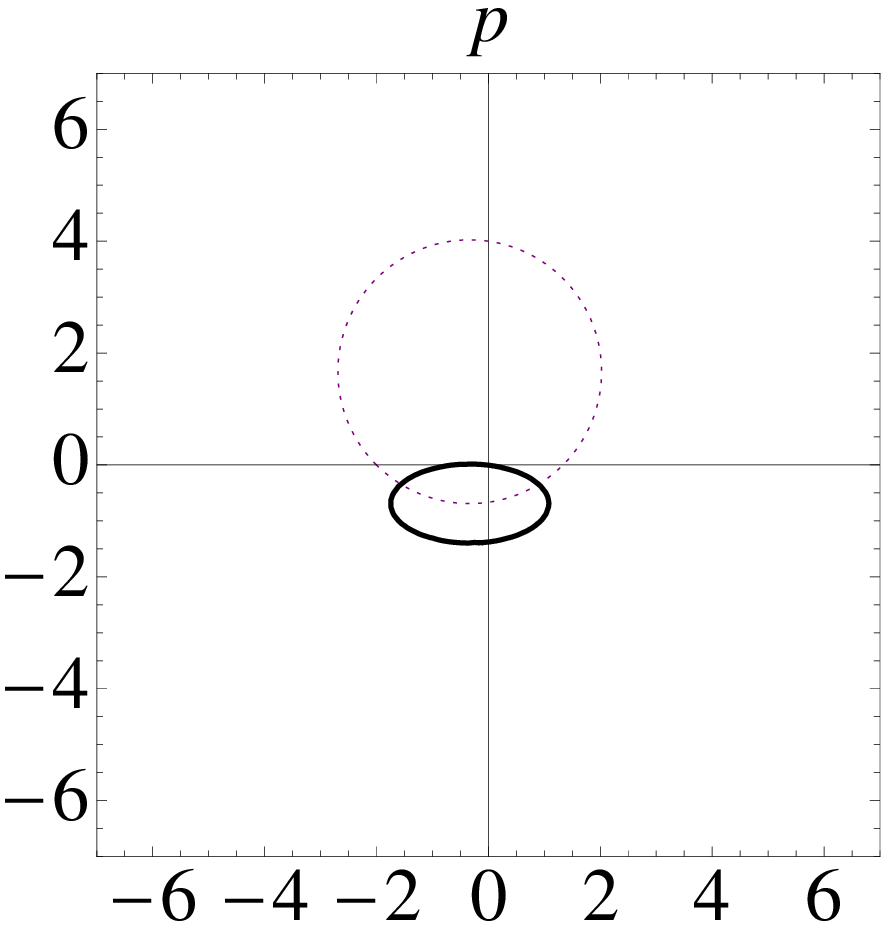}}
\subfloat[$\Om t=\frac{\pi}{2}$]{\label{fig:24}\includegraphics[width=0.313\textwidth]{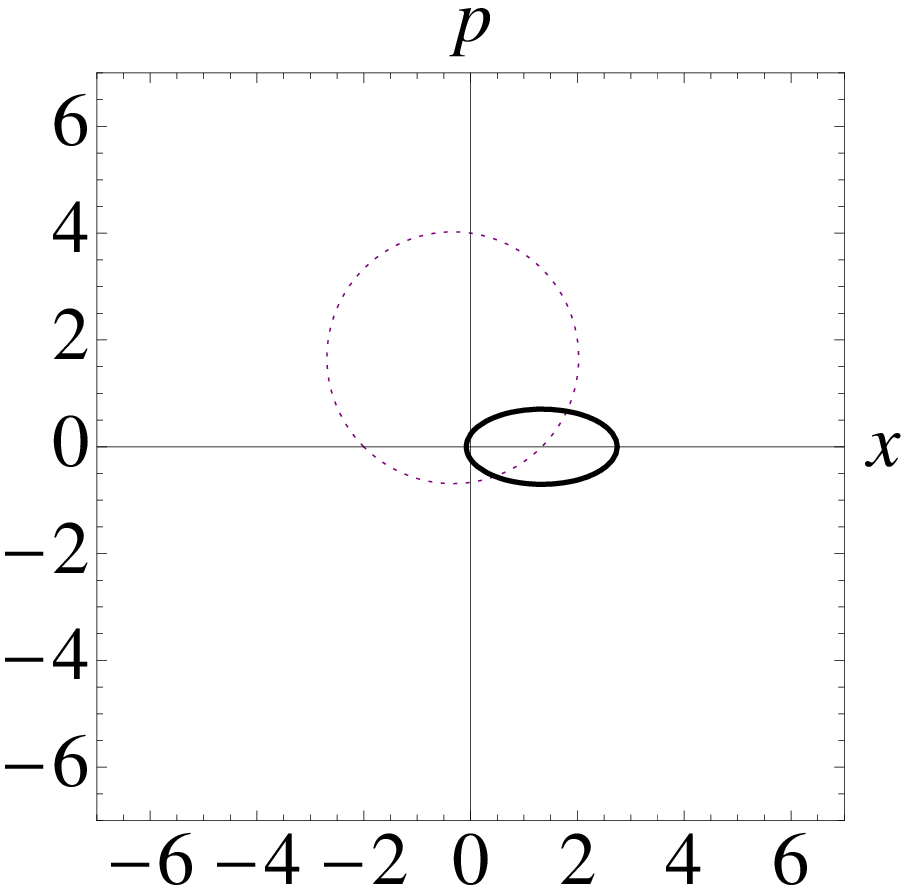}}\\
\subfloat[$\Om t=\frac{3 \pi}{4}$]{\label{fig:25}\includegraphics[width=0.3\textwidth]{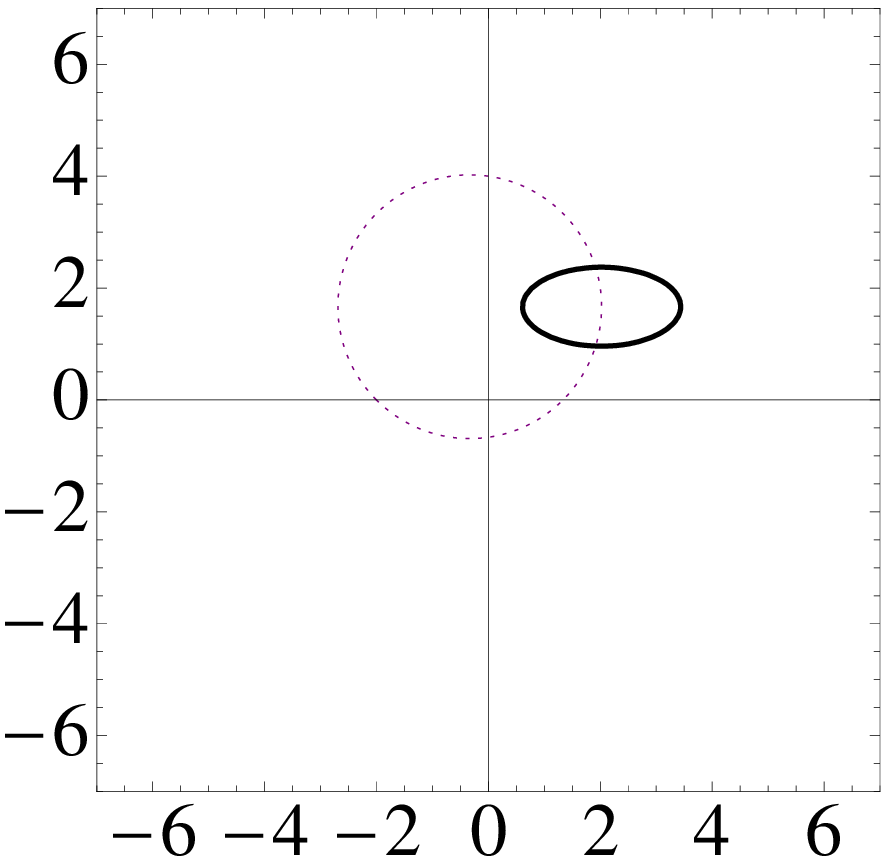}}
\subfloat[$\Om t=\pi$]{\label{fig:26}\includegraphics[width=0.3\textwidth]{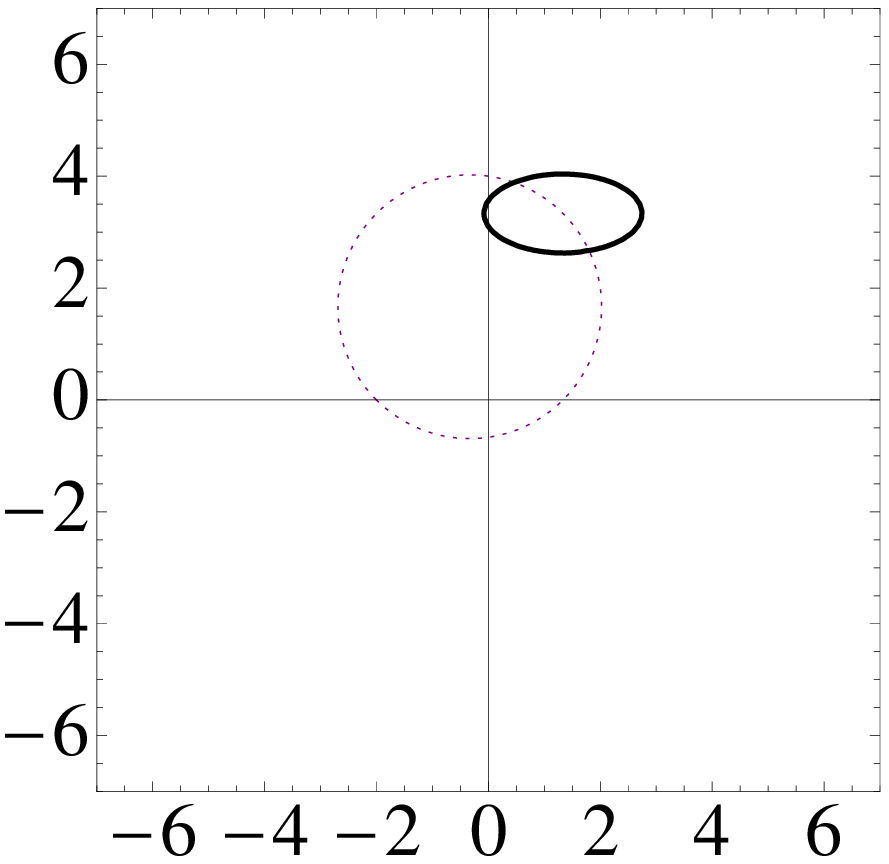}}
\subfloat[$\Om t=\frac{5\pi}{4}$]{\label{fig:27}\includegraphics[width=0.313\textwidth]{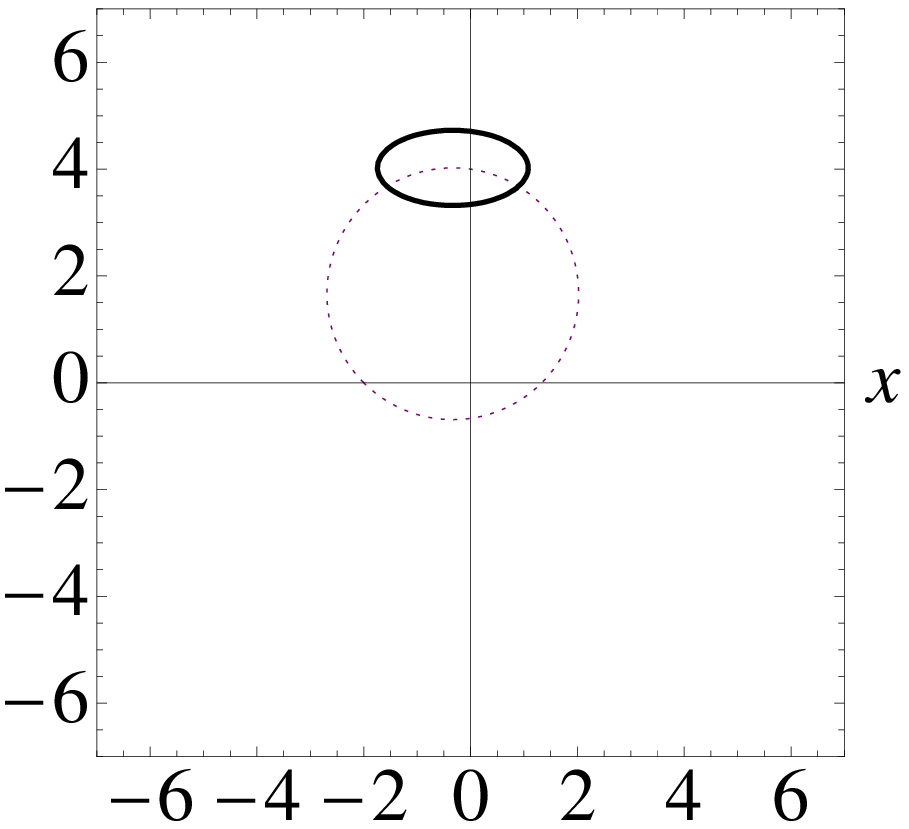}}\\
\subfloat[$\Om t=\frac{3 \pi}{2}$]{\label{fig:28}\includegraphics[width=0.3\textwidth]{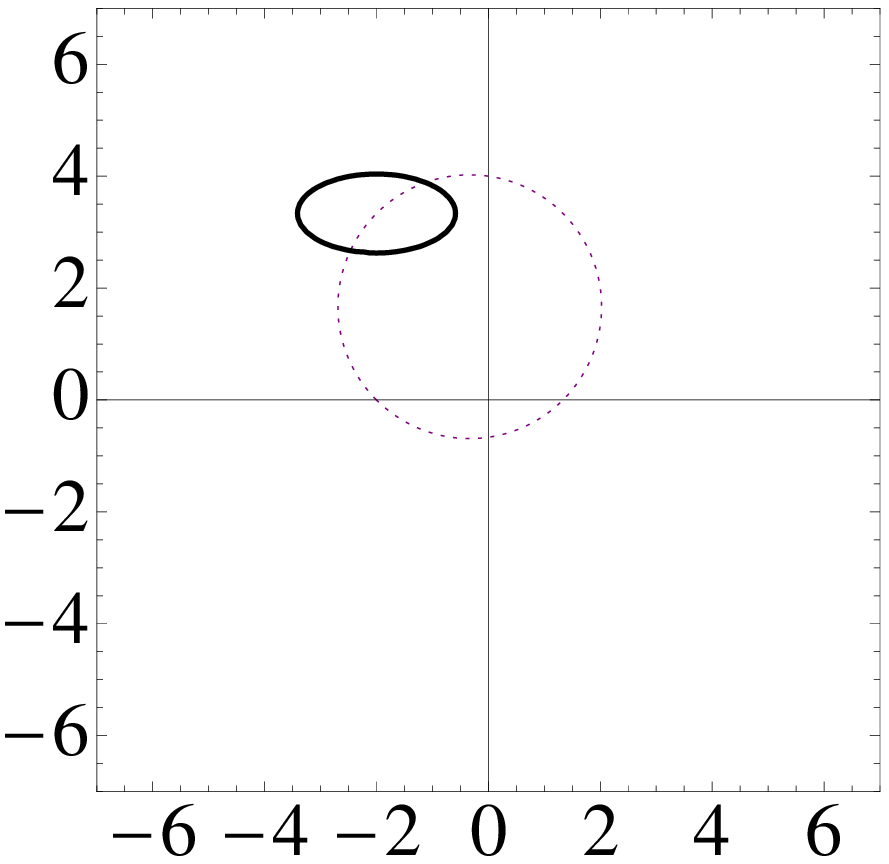}}
\subfloat[$\Om t=\frac{7\pi}{4}$]{\label{fig:29i}\includegraphics[width=0.3\textwidth]{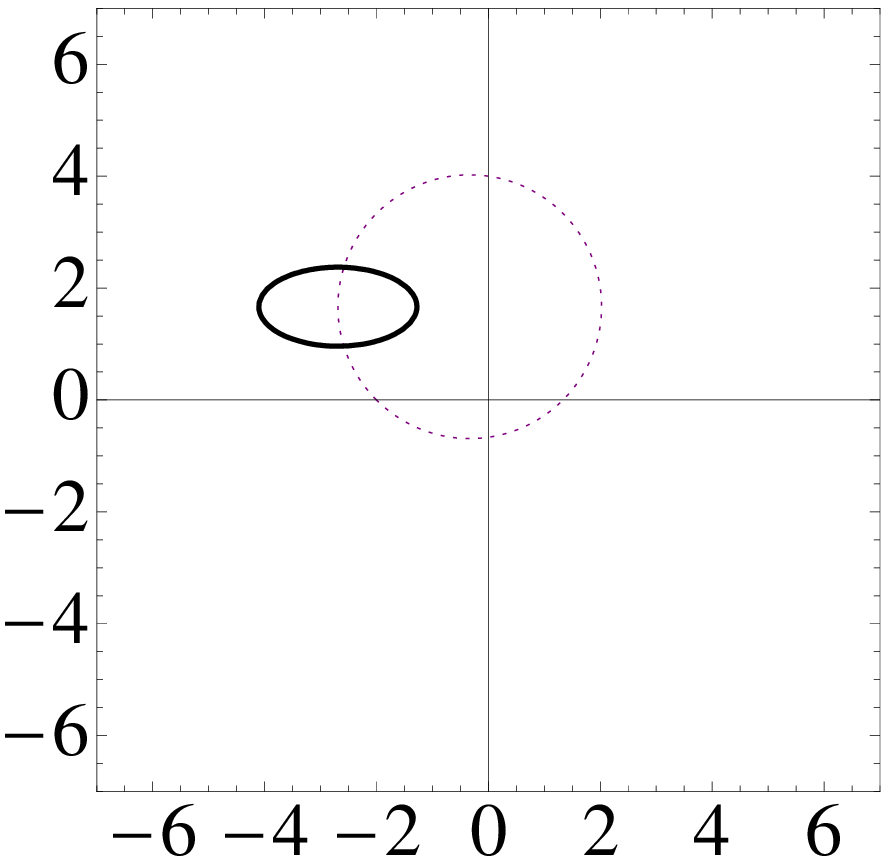}}
\subfloat[$\Om t=2 \pi$]{\label{fig:29j}\includegraphics[width=0.313\textwidth]{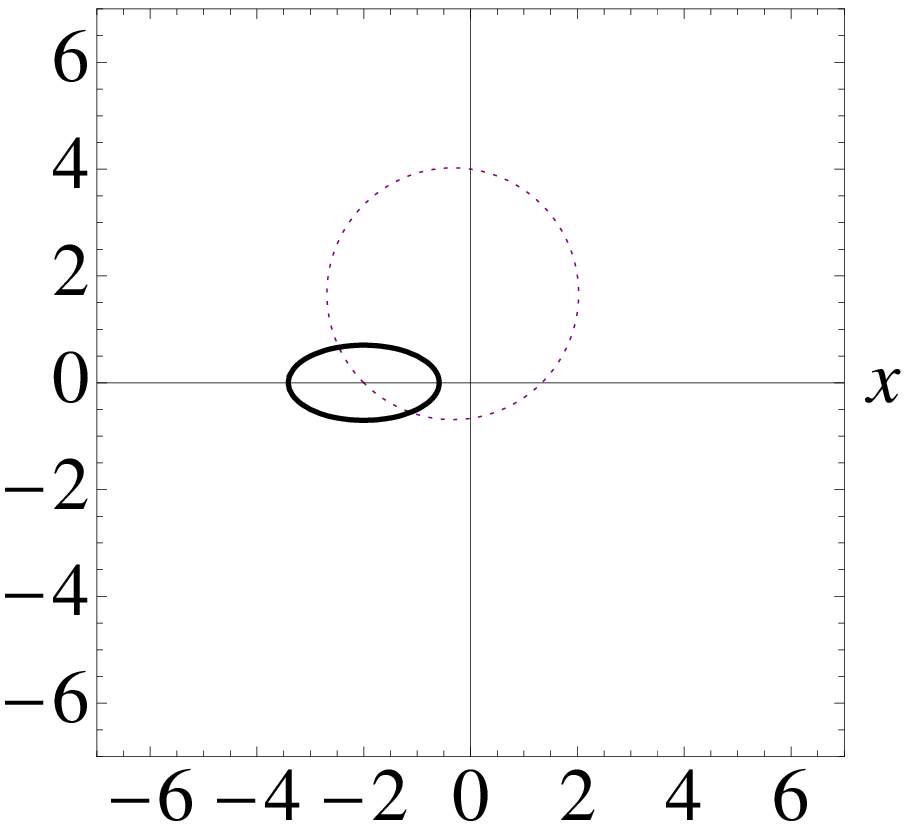}}
\caption{Time-evolution of the $\frac{1}{\rme}$-contour of the IP-Wigner function  ($\sigma_x = 1, \sigma_p = \half, \mu_x = -2, a=1, b= -1, \Om =3, g = 5$) in one period. The evolution corresponds to its global counterclockwise parallel transformation about the origin on the dotted circle described by equation (\ref{breath}).}
\label{fig.2}
\end{figure}

For a system that is initially in an ideal squeezed state described by (\ref{int_ro_ini}), the state at later time $t$ in the interaction picture reads
\begin{equation}\label{int_ro_t}
\hro_{\textnormal I}(t) = \D \left(\nu(t)\right) \D (\frac{g}{\Om} \alpha^*)\hro(0)\D (- \frac{g}{\Om} \alpha^*)\D \left(- \nu(t)\right), 
\end{equation}
where we substituted Eq. (\ref{UI}) for $\ui$ in (\ref{int-w-sol}).
Knowing the state, we can construct the IP-Wigner function following the same line of calculations  as in section~\ref{Sec:Passive} and get
\BEA
W^{\textnormal{I}}_{1}(x,p;t) = \frac{1}{\pi}\exp{\left\{- \half \left[ \bi r - \bi {\bmu_{\textnormal I}}(t) \right]^T \cdot \Xi^{- 1} \cdot \left[ \bi r - \bmui(t) \right]\right\}},
\EEA
where the suffix $\textnormal I$ represents the IP, and the index $1$ represents the evolution of the Wigner function the takes place in the presence of the interaction Hamiltonian $\hH_1(t)$ given by (\ref{H1_def}). The vector $\bmui(t)$ is defined as
\begin{equation}\label{r_int_mean}
\bmui(t)=\left(\begin{array}{c}\mu_{x} \\0\end{array}\right) + \gO \left(\begin{array}{c}a \\- b\end{array}\right) -\gO R(\Om t) \left(\begin{array}{c}a \\- b\end{array}\right).
\end{equation}
In order to see how the Wigner function evolves in the IP, we look at the equation of motion of its centroid given by
\begin{equation}\label{breath}
\left(\langle x_{\textnormal I}(t) \rangle - \mu_x\right)^2 + \left( \langle p_{\textnormal I}(t) \rangle \right)^2 = 2 \left(\gO \right)^2 (a^2 + b^2) (1 - \cos \Om t).
\end{equation}
As it is shown in figure~\ref{fig.2}, the time-evolution of the $\frac{1}{\rme}$-contour of the IP-Wigner function corresponds to a displacement of the centroid of its elliptical distribution in phase space together with a counterclockwise parallel transformation of the distribution on a circle with a time-dependent radius described by equation (\ref{breath}). There are two important points that can be inferred from figure~\ref{fig.2}: ${\it i)}$ the squeezing direction remains invariant during the evolution in the interaction picture, which in our example was in the $x-$direction. ${\it ii)}$ The evolution of the IP-Wigner function corresponds to the displacement of its centroid around a circle as it was the case for the free Hamiltonian evolution of the free evolution of SP-Wigner function of an ideal squeezed state shown in figure \ref{fig.1}. The reason is that the IP-unitary transformation $\ui$ is defined as two successive displacement operators apart from a phase factor rather than a rotation. It is exactly  because of this property that for the Hamiltonian (\ref{interaction}) choosing the IP makes the study of the time-evolution of the state much simpler. We will see in the next part that both of the above mentioned properties get spoiled when we transform back to the SP. 
%

\subsection{Back to the Schr\"{o}dinger picture}\label{Sec:back}

Transforming back to the SP amounts to performing the unitary transformation  
\begin{equation}\label{rho_int_schro}
\hro(t) = \hU\hro_{\textnormal I}(t) \hU^{\dagger},
\end{equation}
which as we showed in section~\ref{Sec:Passive} corresponds to a global and a local  rotations of the corresponding SP-Wigner function through angle $\om_0 t$.
Then the symmetric characteristic function is straightforward to establish. As a result the SP-Wigner function of the system in the presence of the interaction Hamiltonian $\hH_1(t)$ is given by
\BEA\label{backschrowig}
W^{\textnormal{SP}}_1(x,p;t) = \frac{1}{\pi} \exp{\left\{- \half \left[ \bi r - \bi{\bmu}_{\textnormal S}(t)\right]^{T}\cdot \Gamma^{-1}\cdot \left[ \bi r - \bi{\bmu}_{\textnormal S}(t) \right] \right\}}.
\EEA
The $2 \times 2$ matirx $\Gamma^{-1}$ is given by (\ref{Gamma_def}) and the mean value vactor $\bi{\bmu}_{\textnormal S}(t)$ is defined as the clockwise rotation of the initial displacement about the origin through angle $\om_0 t$ plus two types of transformations of the interaction-induced displacement about the origin: a clockwise rotation through angle $\om_0t$ and a counterclockwise rotation through angle $\om_1 t$:
\begin{equation}\label{meanr_def}
\bi{\bmu}_{\textnormal S}(t) = R(- \om_{0}t)\left(\begin{array}{c}\mu_x\\0\end{array}\right) + \gO\left[ R(- \om_0 t) \left(\begin{array}{c}a \\- b\end{array}\right) -  R(\om_1 t) \left(\begin{array}{c}a \\- b\end{array}\right) \right].
\end{equation}

\begin{figure}
  \centering
  \subfloat[$\om_1 = \frac{1}{3}$]{\label{fig:o1_3}\includegraphics[width=0.3\textwidth,clip=]{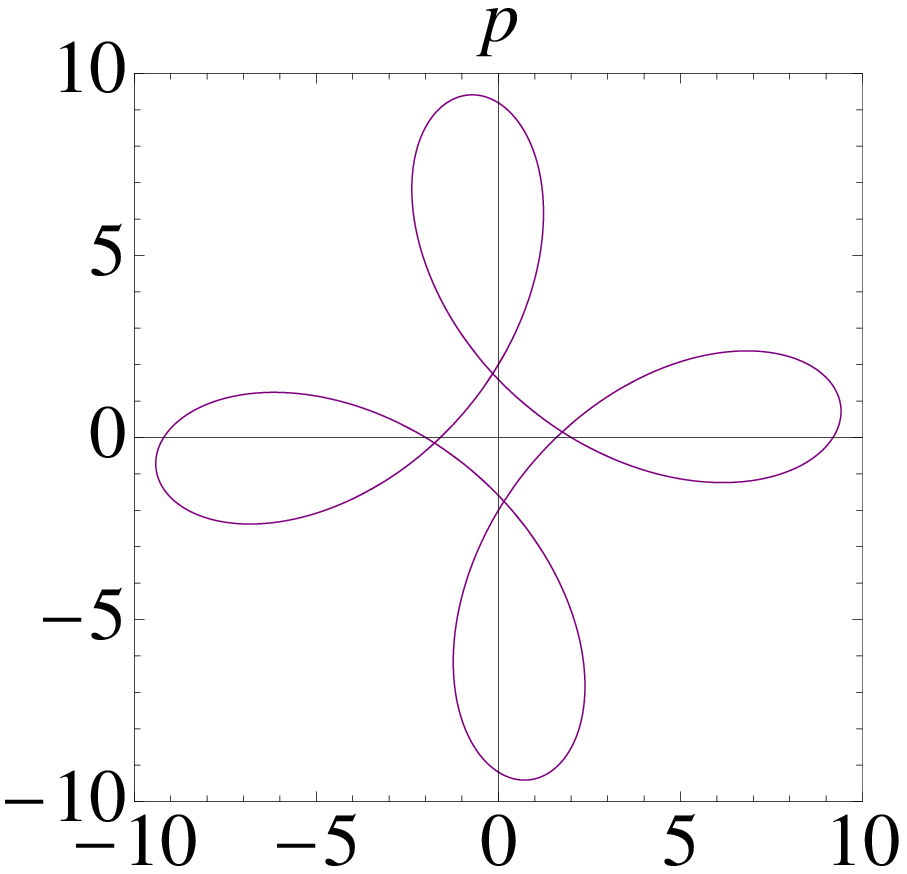}}                
  \subfloat[$\om_1 =  \frac{2}{3}$]{\label{fig:o2_3}\includegraphics[width=0.3\textwidth,clip=]{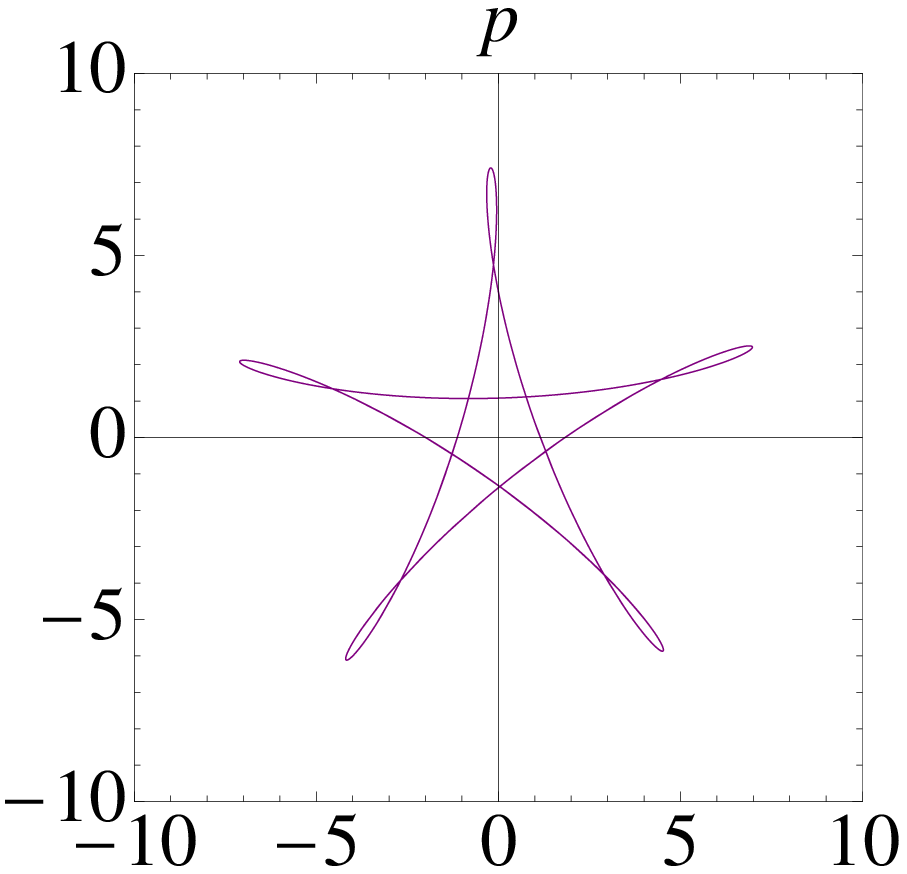}}
  \subfloat[$\om_1 =  \frac{3}{5}$]{\label{fig:o3_5}\includegraphics[width=0.31\textwidth,clip=]{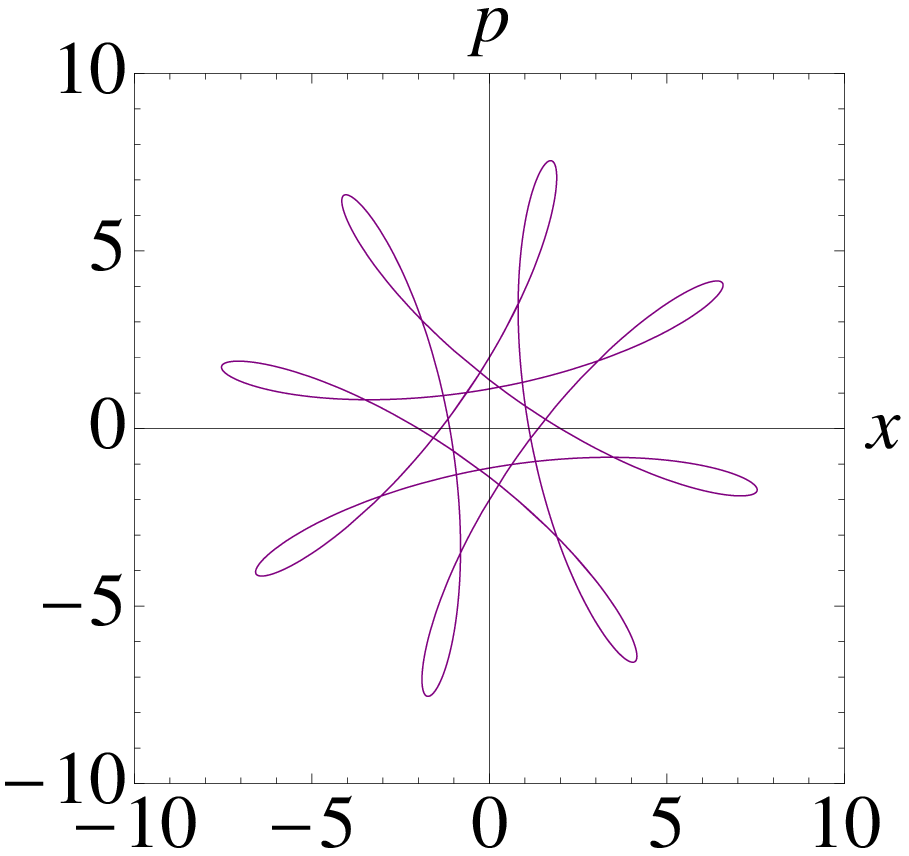}}\\
  \subfloat[$\om_1 = 3$]{\label{fig:o3}\includegraphics[width=0.3\textwidth,clip=]{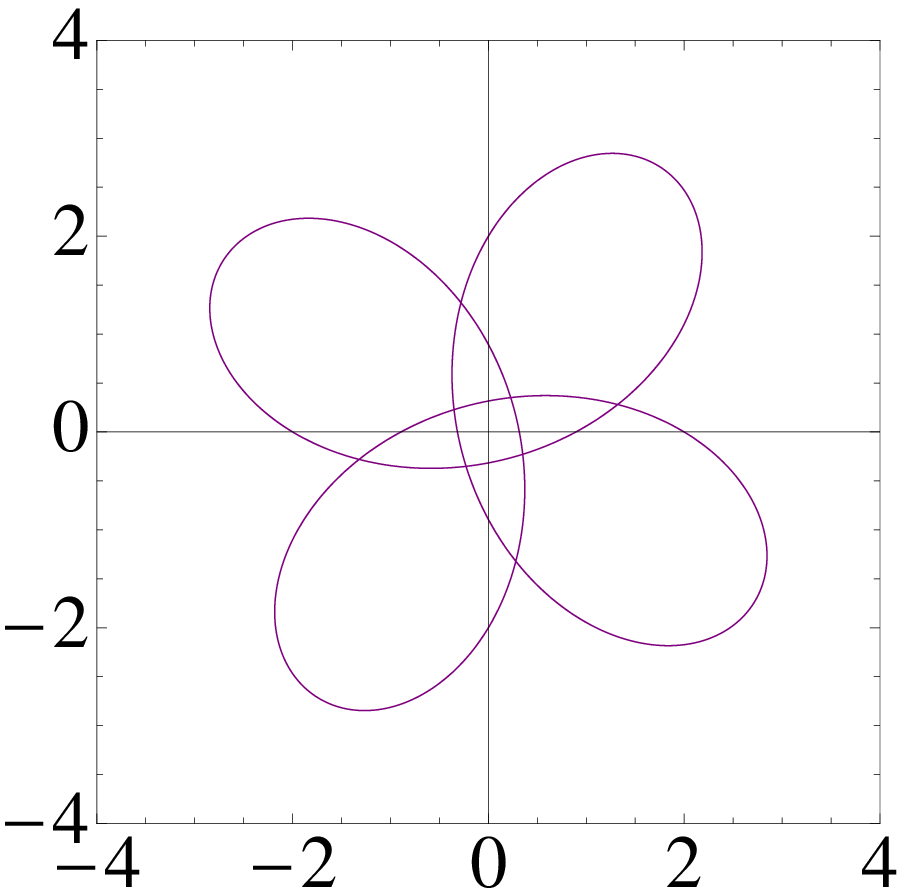}}                
  \subfloat[$\om_1 = 4$]{\label{fig:o4}\includegraphics[width=0.3\textwidth,clip=]{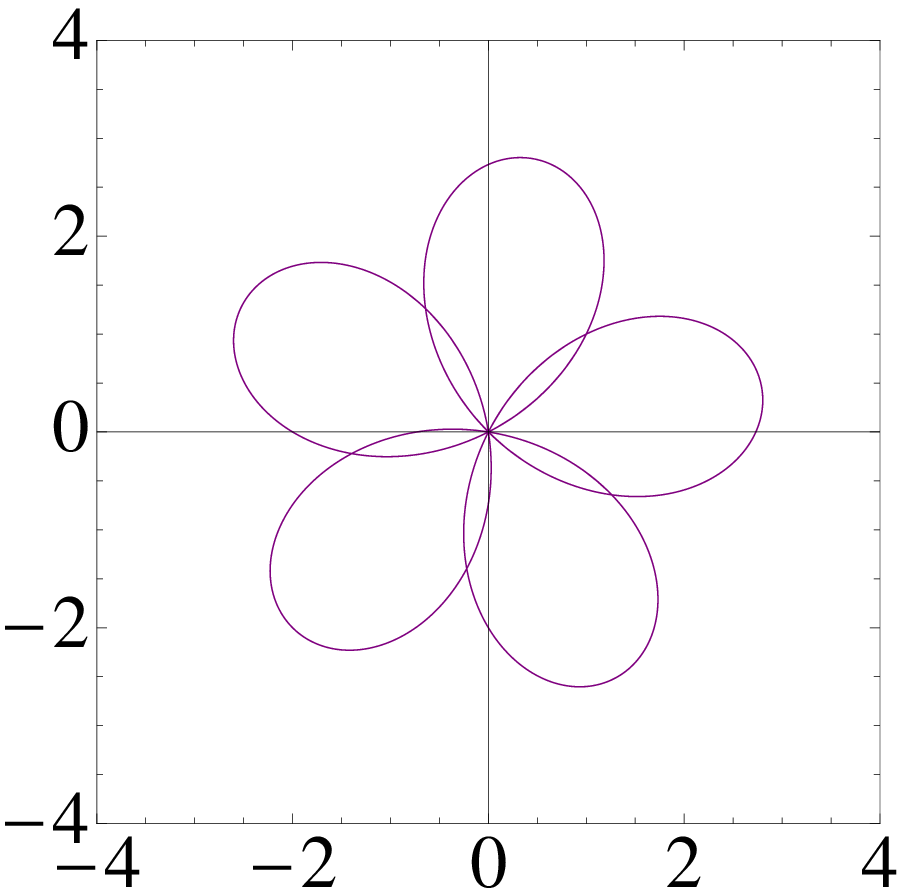}}
  \subfloat[$\om_1 = 4 \frac{5}{7}$]{\label{fig:o33_5}\includegraphics[width=0.313\textwidth,clip=]{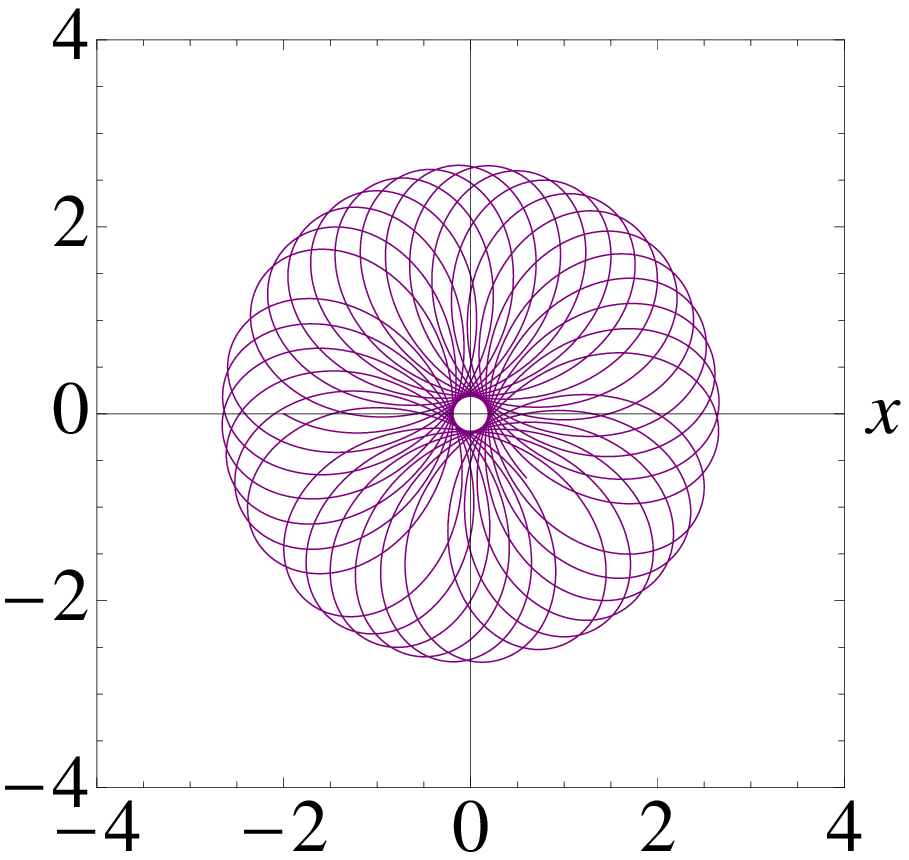}}\\
  \subfloat[$\om_1 = 5$]{\label{fig:o5}\includegraphics[width=0.3\textwidth]{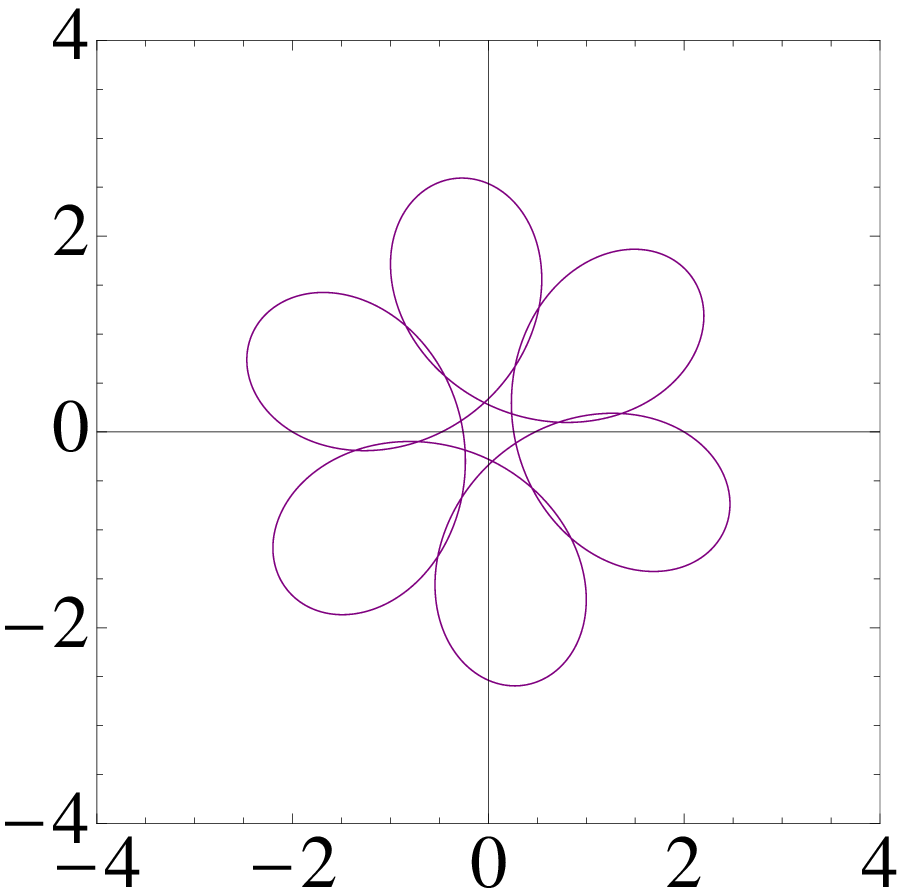}}                
  \subfloat[$\om_1 =  6 \frac{4}{5}$]{\label{fig:o34_5}\includegraphics[width=0.3\textwidth,clip=]{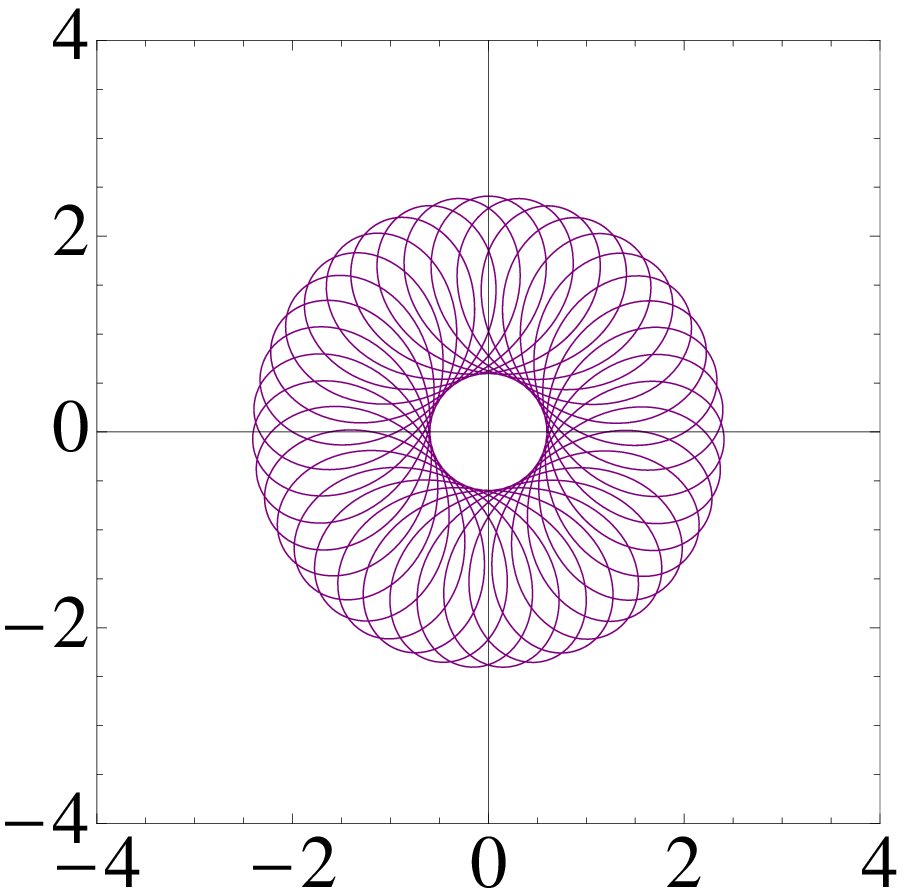}}
  \subfloat[$\om_1 = 9$]{\label{fig:o9}\includegraphics[width=0.313\textwidth,clip=]{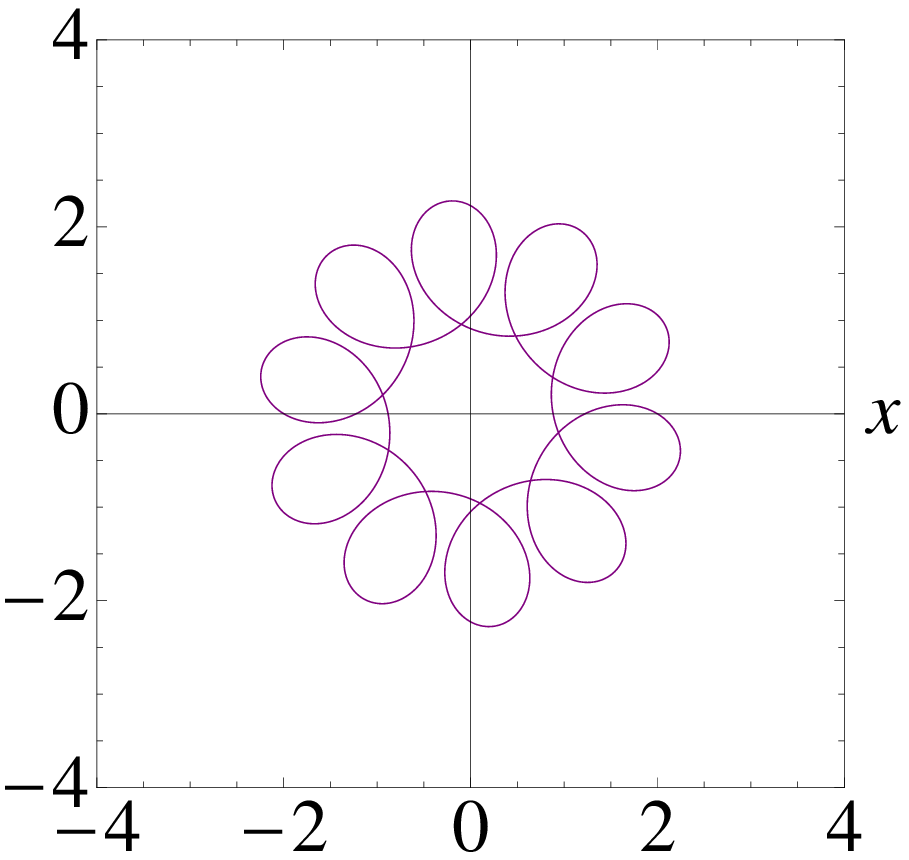}}
  \caption{The trajectory of the center of SP-Wigner function in the presence of the interaction for different values of $\om_1$. $\om_0$ is fixed to $1$ in arbitrary units. The parameters are fixed at $\sigma_x = 1, \sigma_p = \half, \mu_x = -2, a=1, b= -1, g=5$.}
  \label{fig.3}
\end{figure}
The first point we notice by looking at the Wigner function of (\ref{backschrowig}) is that when we go back to the SP the local and global rotations of the Wigner function appear again and this is due to the presence of the unitary transformation $\hU$ . Furthermore, the global rotation of the centroid of the distribution occurs not on a circle, like in the free evolution case illustrated in figure~\ref{fig.2}, but on a more complex trajectory given by
\begin{equation}\label{breathS}
\left(\langle x_{\textnormal S}(t) \rangle - \mu_x \cos \om_0 t\right)^2 + \left( \langle p_{\textnormal S}(t) \rangle + \mu_x \sin \om_0 t \right)^2 = 2 \left(\gO \right)^2 (a^2 + b^2) (1 - \cos \Om t).
\end{equation}
Equation (\ref{breathS}) resembles a glissette~\cite{Roulette}. A glissette is defined as the locus of a generator point which is moving along a given curve. The generator point in (\ref{breathS}) is $(\mu_x \cos \om_0 t, - \mu_x \sin \om_0 t)$ described as the center of the circle in the right hand side of (\ref{breathS}). This point is rolling  along the time-dependent curve given by the left hand side of (\ref{breathS}). The resulting glissette is shown in figure~\ref{fig.3} for different values of $\Om$.
\begin{figure}
\subfloat[$\om_1 t = 0$]{\label{fig:39}\includegraphics[width=0.3\textwidth]{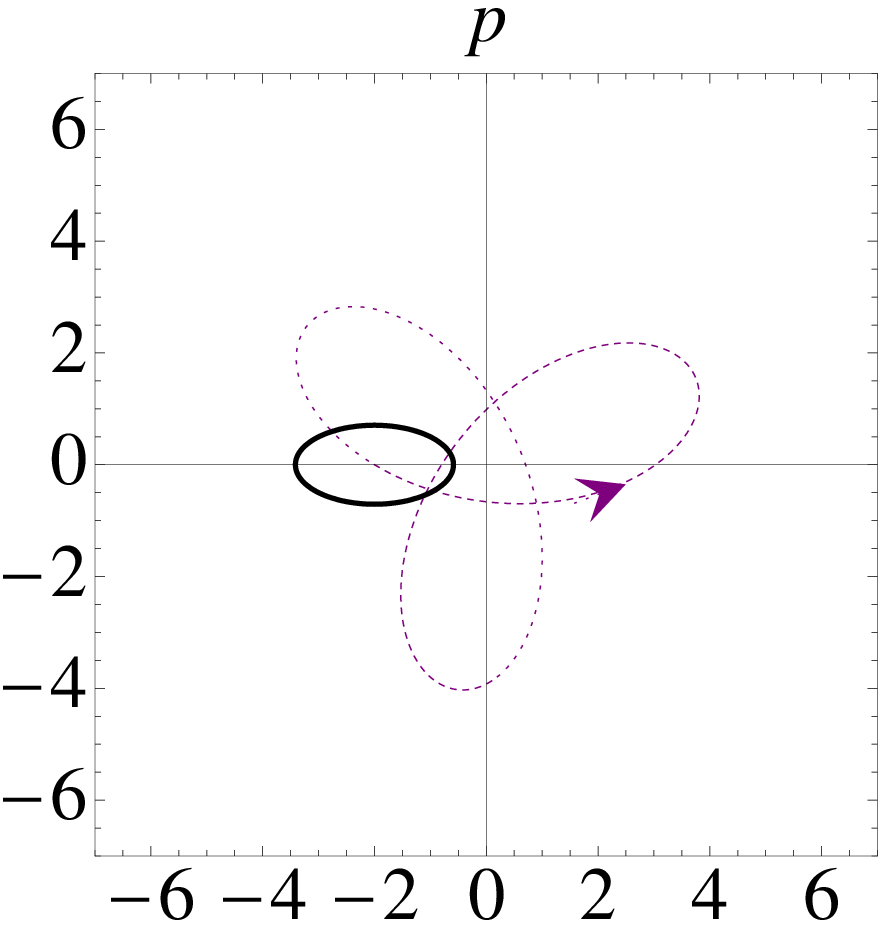}}
\subfloat[$\om_1 t = \frac{\pi}{4}$]{\label{fig:40}\includegraphics[width=0.3\textwidth]{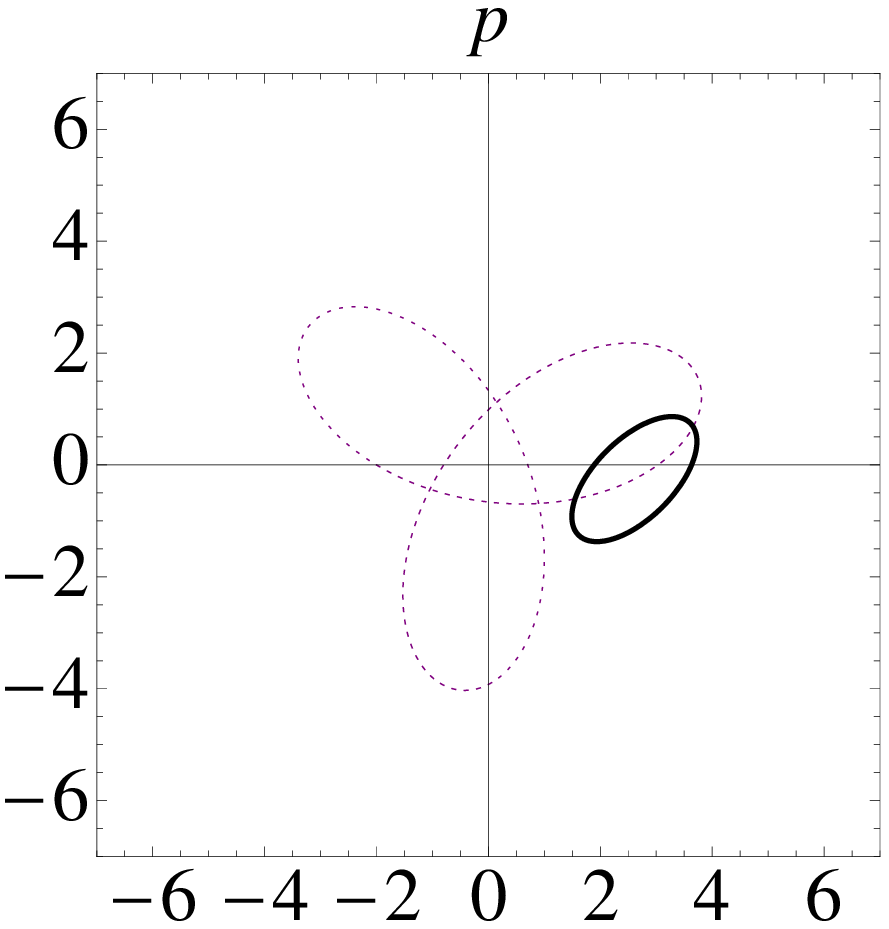}}
\subfloat[$\om_1 t = \frac{\pi}{2}$]{\label{fig:41}\includegraphics[width=0.313\textwidth]{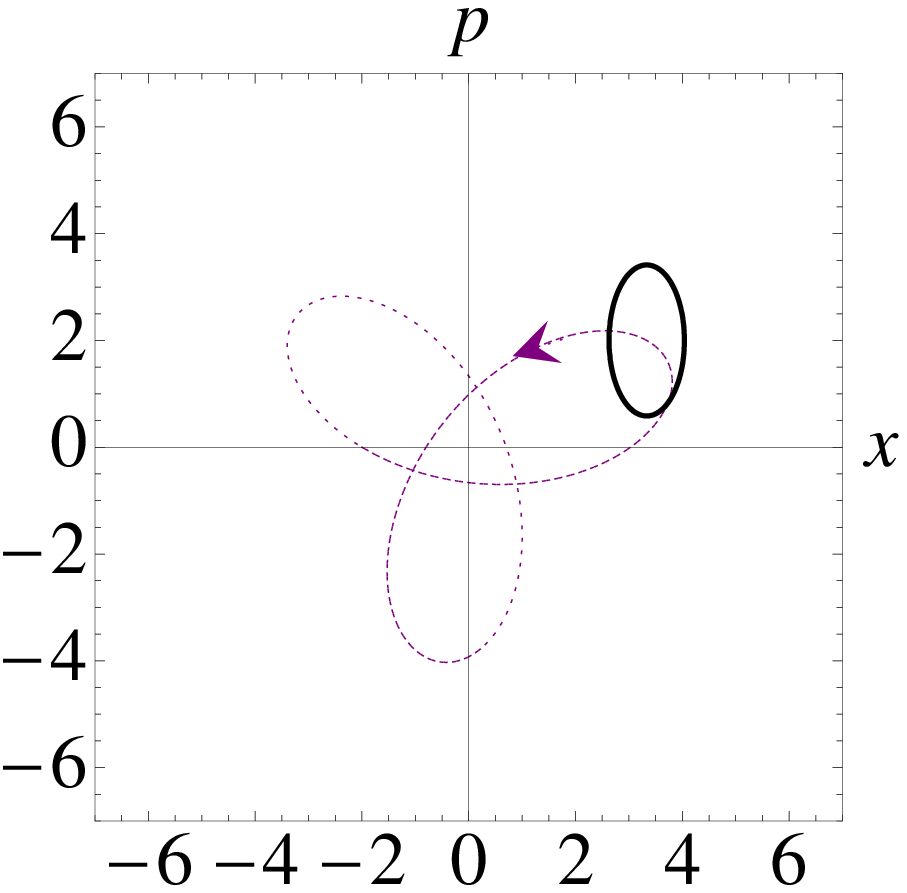}}\\
\subfloat[$\om_1 t= \frac{3 \pi}{4}$]{\label{fig:42}\includegraphics[width=0.3\textwidth]{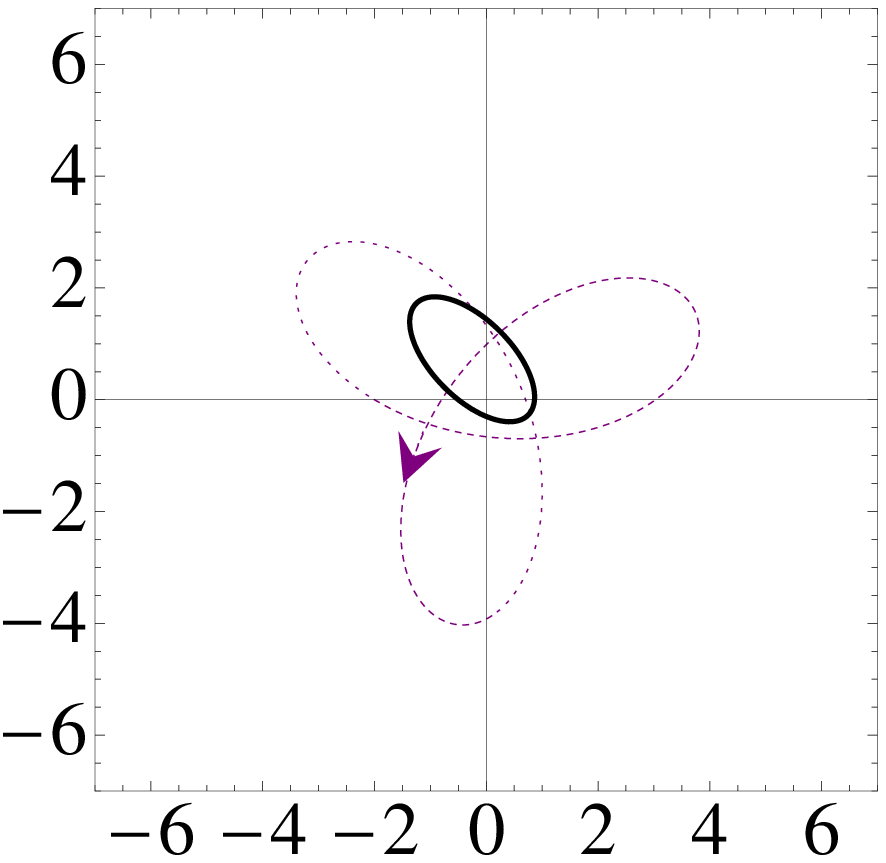}}
\subfloat[$\om_1 t = \pi$]{\label{fig:43}\includegraphics[width=0.3\textwidth]{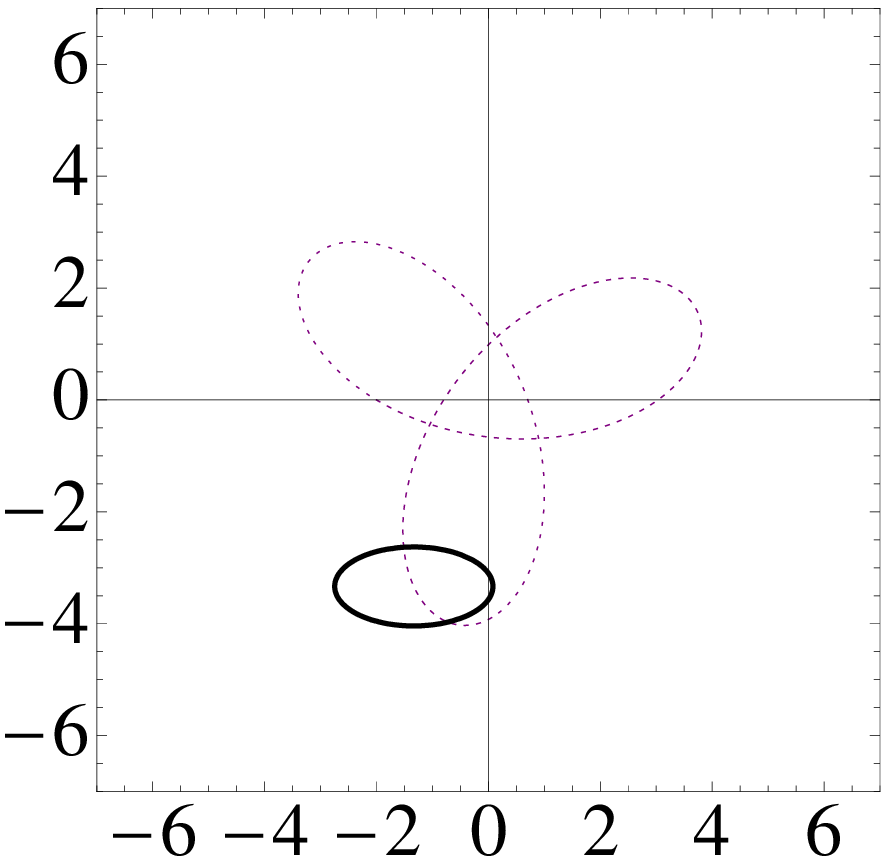}}
\subfloat[$\om_1 t = \frac{5 \pi}{4}$]{\label{fig:44}\includegraphics[width=0.313\textwidth]{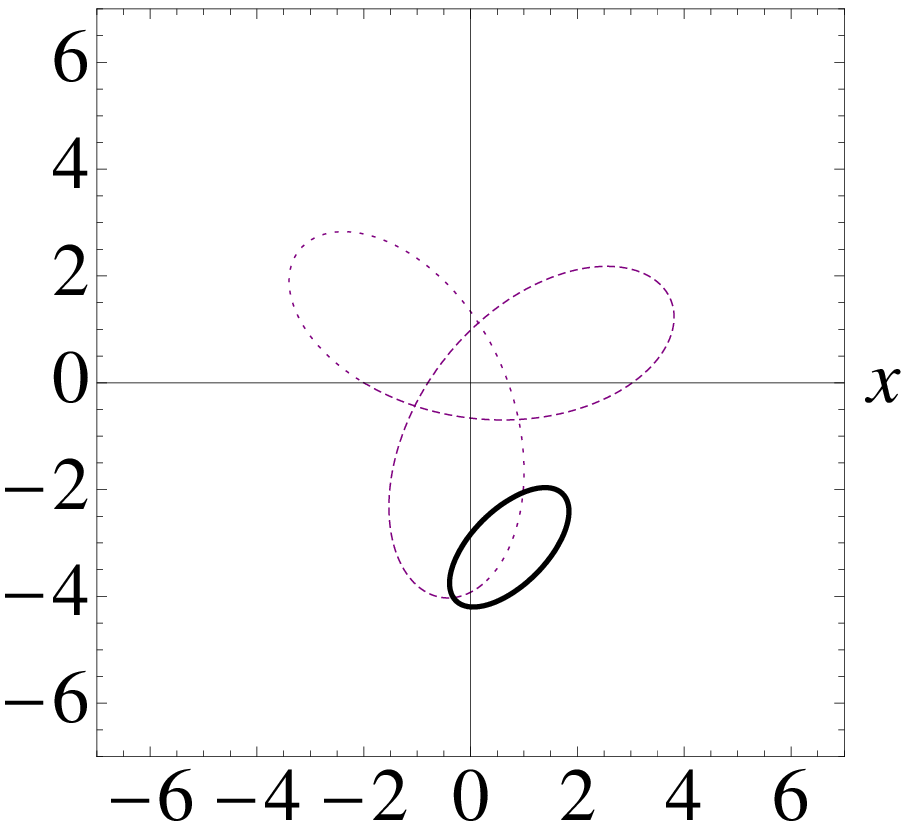}}\\
\subfloat[$\om_1 t = \frac{3 \pi}{2}$]{\label{fig:45}\includegraphics[width=0.3\textwidth]{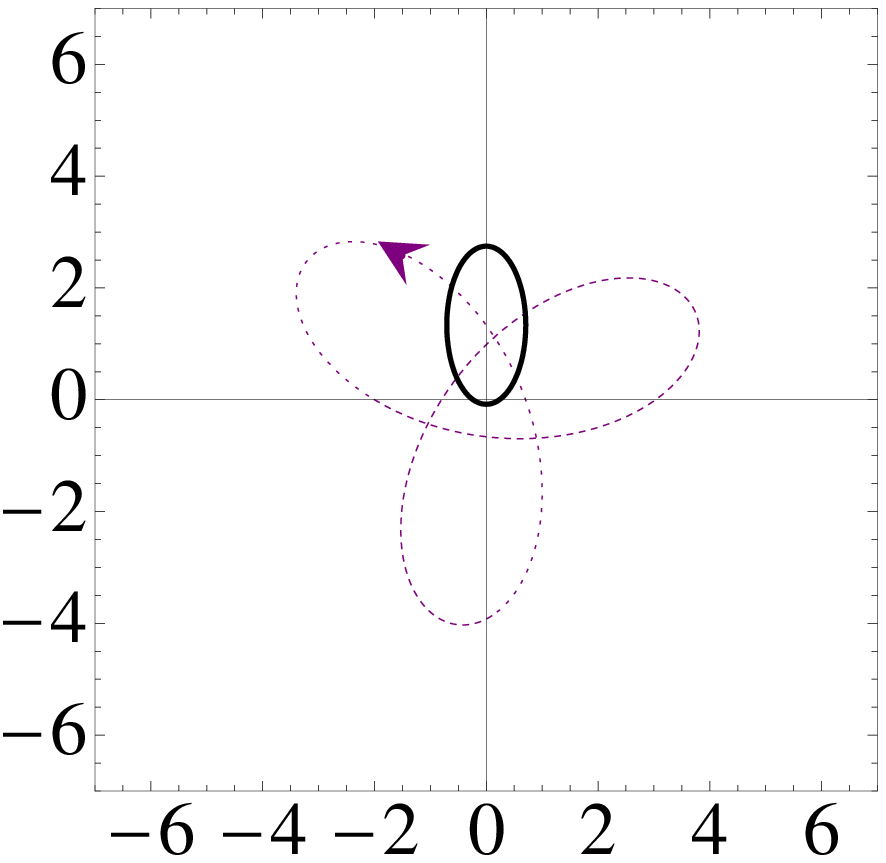}}
\subfloat[$\om_1 t = \frac{7 \pi}{4}$]{\label{fig:46}\includegraphics[width=0.3\textwidth]{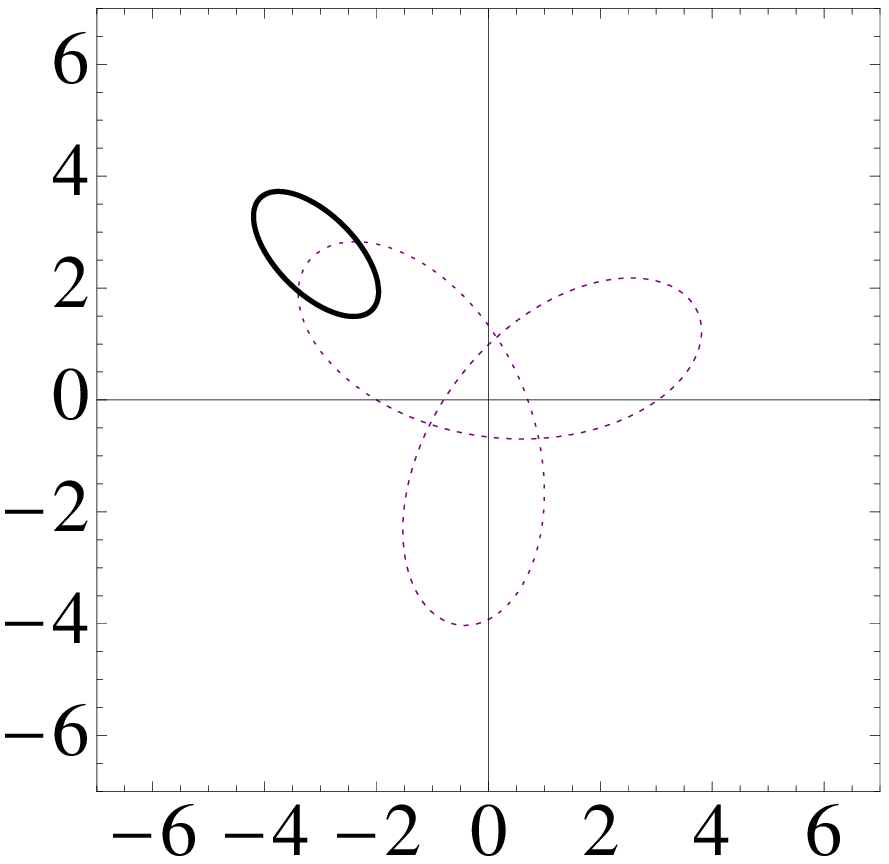}}
\subfloat[$\om_1 t = 2 \pi$]{\label{fig:47}\includegraphics[width=0.313\textwidth]{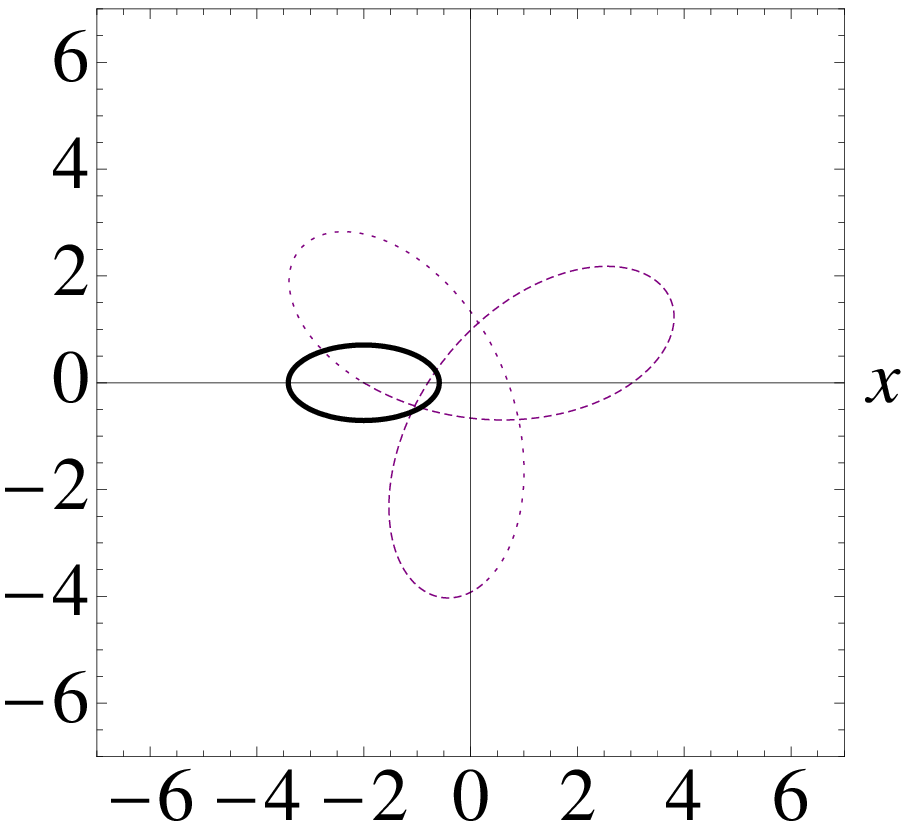}}
\caption{Time-evolution of the $\frac{1}{\rme}$-contour of $W^{\textnormal SP}_{1}(x,p;t)$  given by (\ref{backschrowig}) ($\sigma_x = 1, \sigma_p = \half, \mu_x = -2, a=1, b= -1, \om_1 = 2, \om_0 =1, g = 5$). The evolution is a combination of local rotations of the center of the ellipse through two different angles $\om_{0}t$ and $\om_{1}t$ and a global rotation of the ellipse through angle $\om_{0}t$. This leads to  the ``dancing'' of $W^{\textnormal{SP}}_{1}(x,p;t)$. The dotted curve describes the motion of the centroid given by (\ref{breathS}).}
\label{fig.4}
\end{figure}

On top of the rotations induced by $\hU$, there are two more rotations of the driven displacement given by $\alpha^{*}$ due to the presence of the interaction. ${\it i)}$ Its clockwise rotation through angle $\om_{0}t$; ${\it ii)}$ its counterclockwise rotation through angle $\om_1t$. As a result, the SP-Wigner function in the presence of the interaction ``dances" in a complicated manner, in comparison with its parallel transform in IP, in phase space. This is illustrated  in figure~\ref{fig.4} for a fixed value of $\Om$. A comparison between the figure \ref{fig.2} and figure \ref{fig.4} makes it clear why in studying the evolution of the state of our system of interest the interaction picture is preferred.

Choosing the counterclockwise rotation of the $xp$-plane given by (\ref{r-r'}) transforms the SP-Wigner function into the IP-Wigner function in the local $x'p'$ frame. This corresponds to performing the unitary transformation $\hU$ on the observables in Hilbert space. Correspondingly, in phase space replacing $\bi r$ in (\ref{backschrowig}) with $R(- \om_0\,t)\,\bi{r'}$ yields the same result. Then the vector $\bi{r} - \bi{\bmu}_{\textnormal S}(t)$ in the local reference frame reads 
\BEA\label{rminmus_new}
\fl\bi{r} - \bi{\bmu}_{\textnormal S}(t) \rightarrow R(- \om_0\, t) \left(\bi{r'} - \bi{\bmu}\right) - \gO R(- \om_0\, t) \left(\begin{array}{c}a \\- b\end{array}\right) + \gO R(\om_1 t) \left(\begin{array}{c}a \\- b\end{array}\right).
\EEA
Substituting (\ref{rminmus_new}) into the expression for the SP-Wigner function given by (\ref{backschrowig}) and taking into account that $\Gamma^{-1}$ is defined as $R(-\om_0\,t) \Xi^{-1} R(\om_0\,t)$, we get
\BEA\label{backshro_int_new}
\fl W^{\textnormal{SP}}_{1}(x,p;t) \rightarrow W^{\textnormal{I}}_{1}(x',p';t) = \frac{1}{\pi} \exp{\left[- \half \left( \bi{r'} - \bi{\bmu}_{\textnormal I}(t)\right)^{T}\cdot \Xi^{-1}\cdot \left( \bi{r'} - \bi{\bmu}_{\textnormal I}(t) \right) \right]}.
\EEA
In fact, by choosing the counterclockwise rotation of the coordinates, we let the observables of the system evolve under the free Hamiltonian $\hH_0$. Thus the evolution of the state in the new local frame is only governed by the interaction Hamiltonian. In other words, we separate the time-evolution of the state from that of the observables. But this is exactly the definition of the Schr\"{o}dinger interaction picture~\cite{SHIP}.

We can also transform the Wigner function to the Heisenberg interaction picture~\cite{SHIP}. Such transformation may be performed by ${\it i)}$ a counterclockwise rotation of $xp$ frame, i.e. a passive transformation of $xp$-frame to $x'p'$-frame; ${\it ii)}$ displacing the origin of $x'p'$-frame by $\gO \tiny{\left(\begin{array}{c}a \\ - b\end{array}\right)}$; ${\it iii)}$ counterclockwise rotation of the new origin of the $x'p'$-frame with the frequency $\Om$. Thus the transformation from $xp$-plane to the new $x''p''$-plane is defined as
\BEA\label{r-r''}
\bi r \rightarrow \bi{r''} =  R(\om_0\,t) \bi r - \gO \left(\begin{array}{c}a \\- b\end{array}\right) + \gO R(\Om\,t) \left(\begin{array}{c}a \\- b\end{array}\right).
\EEA
The last two transformations are precisely compensating the unitary transformation arising from the interaction Hamiltonian. 
It is then straightforward to describe the vector $\bi r - \bi{\bmu}_{\textnormal S}(t)$  in the $x''p''$-frame as the clockwise rotations of $(\bi{r''} - \bmu)$ through angle $\om_0 t$:
\begin{equation}\label{mufree}
\bi r -\bi{\bmu}_{\textnormal S}(t) \rightarrow R(- \om_0 t) (\bi{r''} - \bmu).
\end{equation}
Inserting (\ref{mufree}) into (\ref{backschrowig}) gives us the Wigner function in the local $x''p''$-plane as
\begin{equation}\label{W''}
W^{\textnormal{SP}}_{1}(x,p;t) \rightarrow W^{\textnormal{SP}}_0(x'',p'') = \frac{1}{\pi} \exp{\left[ - \half \left ( \bi{r''} - \bmu \right)^T \cdot \Xi^{-1} \cdot \left( \bi{r''}- \bi{\bmu}\right) \right]}.
\end{equation}
We notice that in the local $x''p''-$frame, only $x''$ and $p''$ are evolving in time according to $\hH(t)$ and the state does not explicitly depend on time.

\subsection{The case $\Om = 0$}\label{sec:Omzero}

It is interesting to see what happens when the driving frequency of the interaction Hamiltonian given in equation (\ref{H1_def}) is equal to that of $\hH_{0}$ with opposite sing. The first thing we notice is that the interaction picture potential given by equation (\ref{VI}) is independent of time. As a result the Wigner function is described by
\BEA\label{Wigne_int_Om0}
W^{\textnormal{I}}_{\Om = 0}(x,p;t) = \frac{1}{\pi} \exp{\left[- \half \left( \bi r - \bmui(t)\vert_{\Om=0}\right)^{T}\cdot\Xi \cdot \left( \bi r - \bmui(t)\vert_{\Om=0}\right) \right]},
\EEA
where the components of $\bmui(t)\vert_{\Om=0}$ are given by
\numparts
\BEA\label{xmean_int_Om0}
\langle x_{\textnormal I}(t) \rangle\vert_{\Om = 0}& = \mu_{x} - g b t,\\
\label{pmean_int_Om0}
\langle p_{\textnormal I}(t) \rangle\vert_{\Om = 0} &= - g a t.
\EEA
\endnumparts
\begin{figure}
\centering
 \includegraphics[width=0.3\textwidth,clip=]{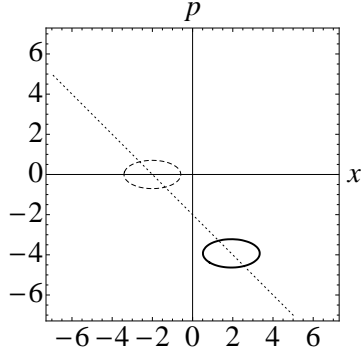}
 \caption{A snapshot of the interaction picture Wigner function $W^{\textnormal{I}}_{1}(x,p;t)\vert_{\Om = 0}$ at some time $t$ with the parameters $\sigma_x = 1, \sigma_p = \half,\mu_x = -2, a=1, b= -1, \om_1 = - \om_0 , g = 5$. The dotted ellipse represent the Winger function at time $t = 0$. The dashed line is the trajectory of the center of the ellipse.}
\label{fig.5}
\end{figure}
Thus the equation of motion for the centroid of the ellipse will be a line instead of a circle and the Wigner function will only be parallel displaced along this line in time. The displacement takes place on a line with slope $a/b$, as it is shown in figure~\ref{fig.5}, and is given by the following equation
\begin{equation}\label{meanx,meanp_int_Om0}
\langle p_{\textnormal I}(t) \rangle\vert_{\Om = 0} = \frac{a}{b} \left( \langle x_{\textnormal I}(t) \rangle\vert_{\Om = 0} - \mu_{x} \right).
\end{equation}

\subsection{Quadratic Hamiltonian}\label{sec:qua}

\begin{figure}
\subfloat[$t = 0$]{\label{fig:70}\includegraphics[width=0.3\textwidth]{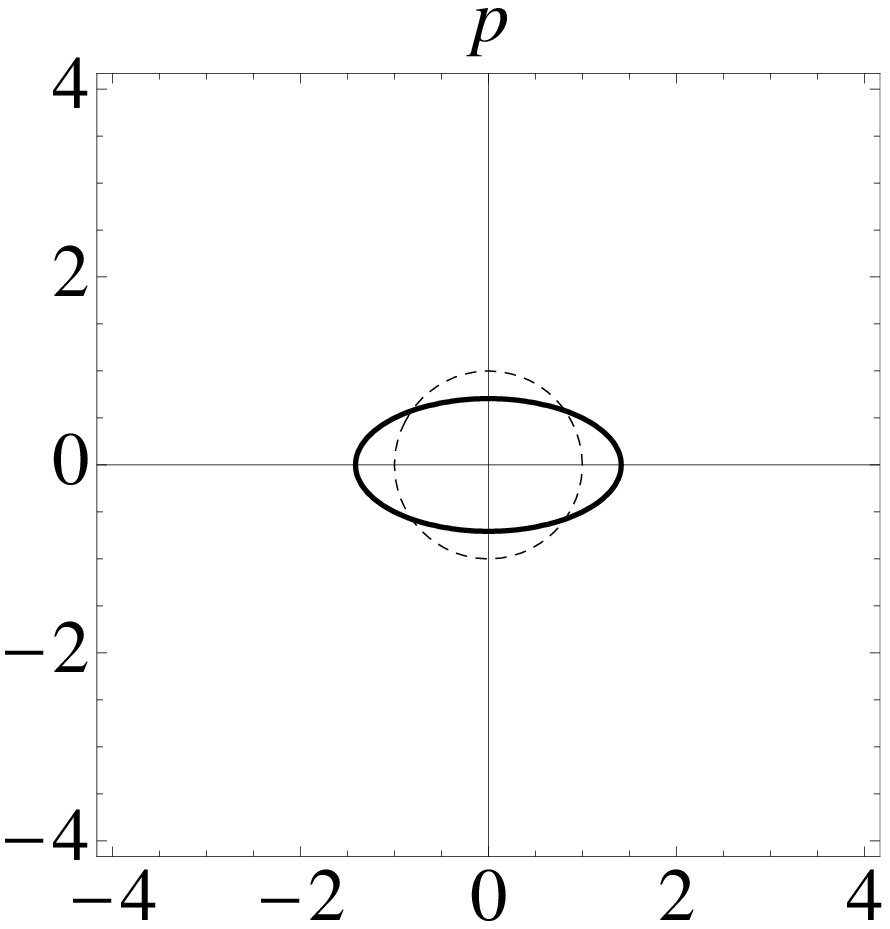}}
\subfloat[$ t = \frac{\pi}{4}$]{\label{fig:71}\includegraphics[width=0.3\textwidth]{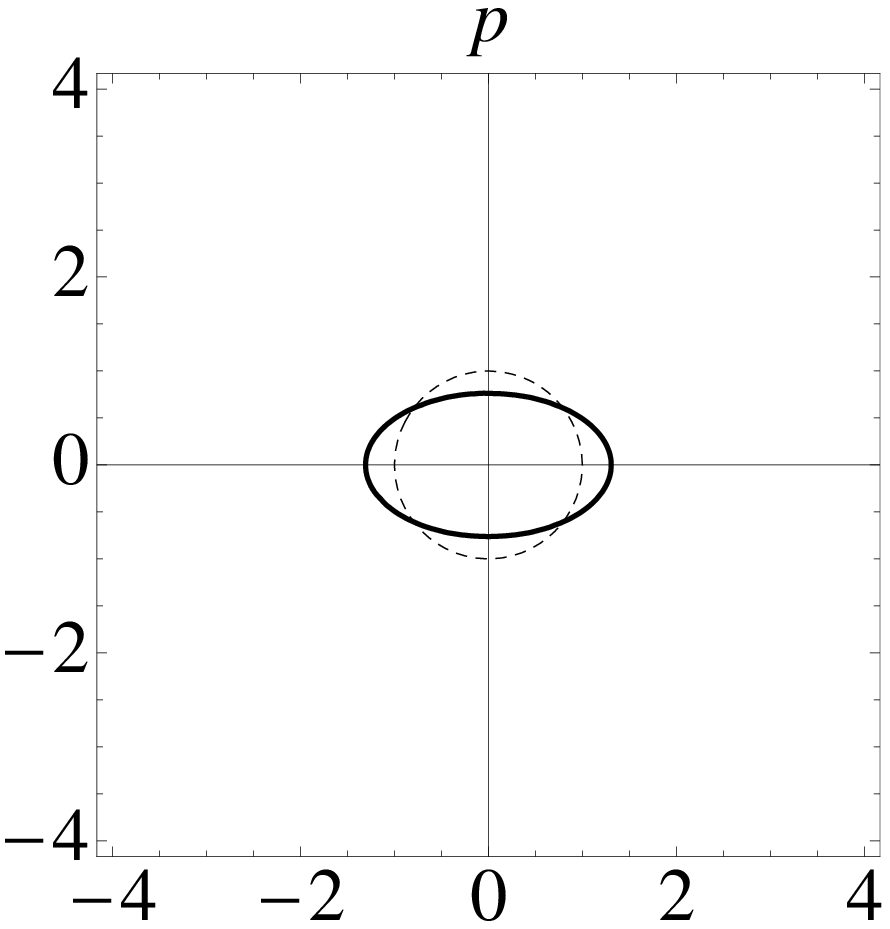}}
\subfloat[$t = \frac{\pi}{2}$]{\label{fig:72}\includegraphics[width=0.313\textwidth]{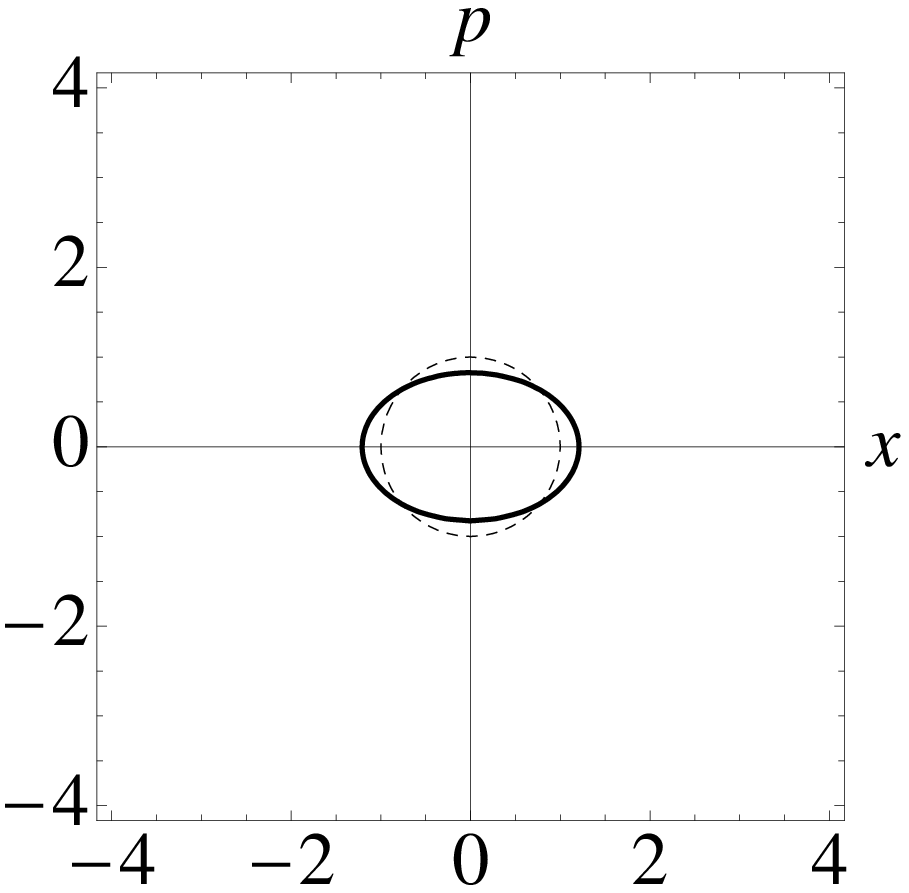}}\\
\subfloat[$t= \frac{3 \pi}{4}$]{\label{fig:73}\includegraphics[width=0.3\textwidth]{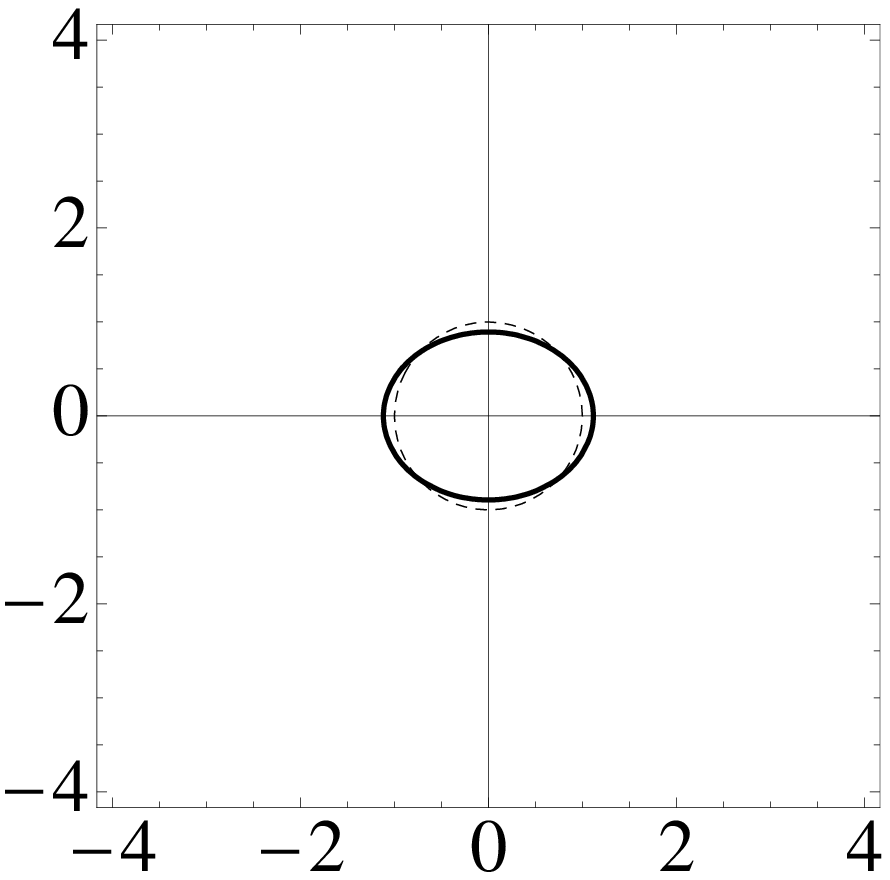}}
\subfloat[$t = \pi$]{\label{fig:74}\includegraphics[width=0.3\textwidth]{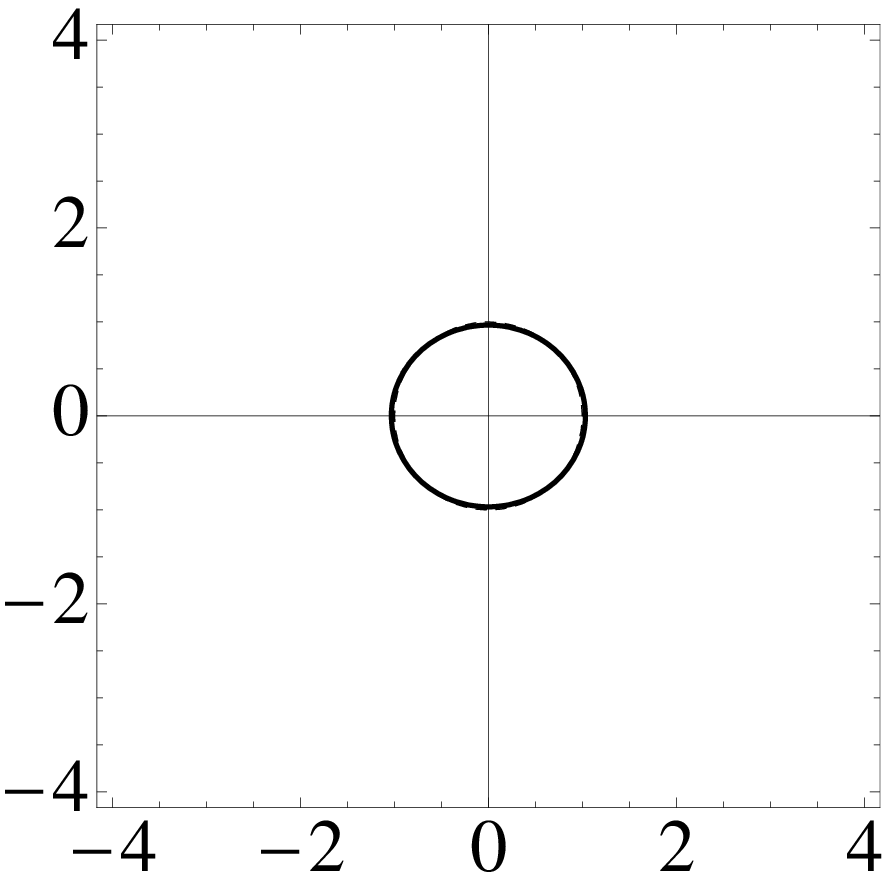}}
\subfloat[$ t = \frac{5 \pi}{4}$]{\label{fig:75}\includegraphics[width=0.313\textwidth]{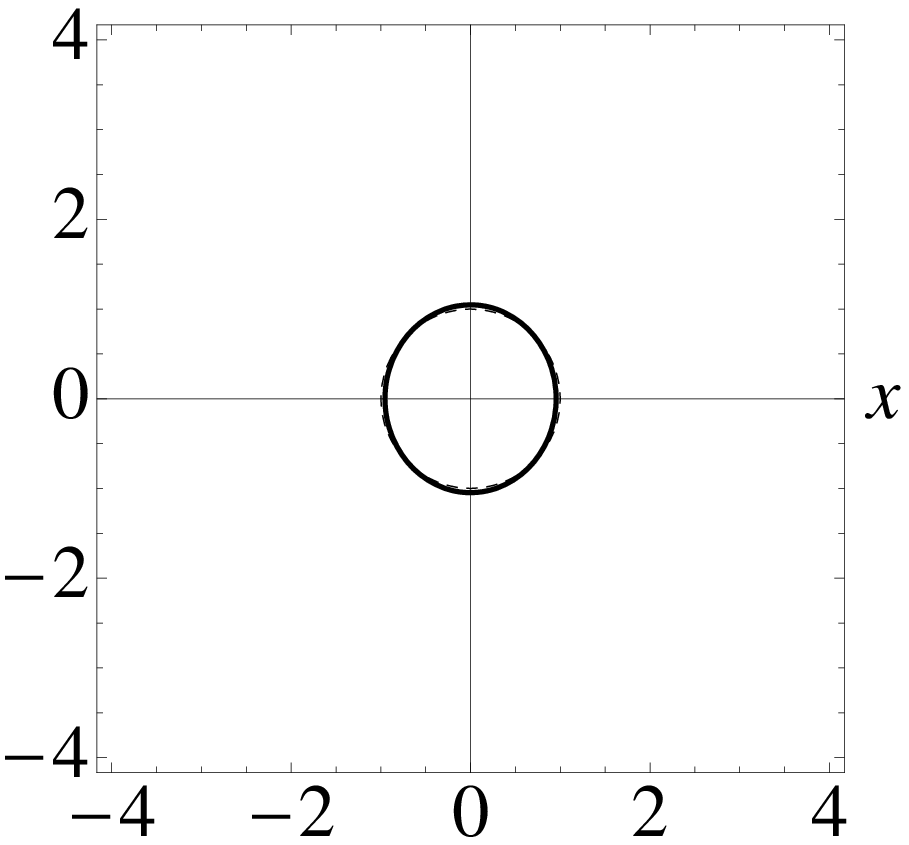}}\\
\subfloat[$t = \frac{3 \pi}{2}$]{\label{fig:76}\includegraphics[width=0.3\textwidth]{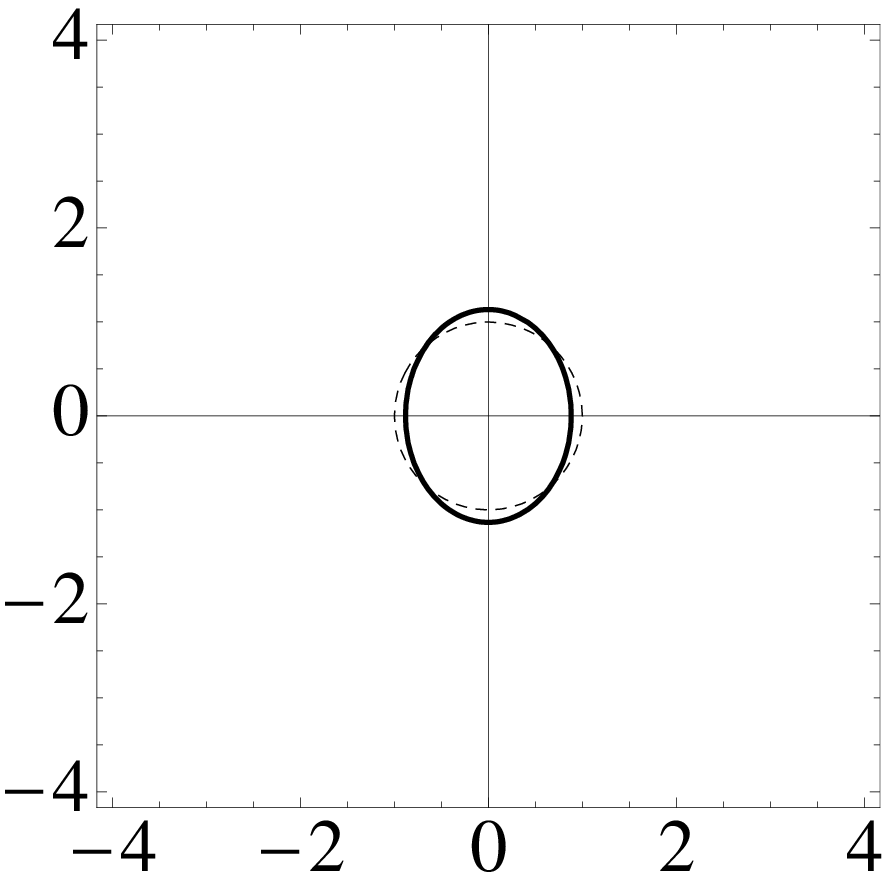}}
\subfloat[$t = \frac{7 \pi}{4}$]{\label{fig:77}\includegraphics[width=0.3\textwidth]{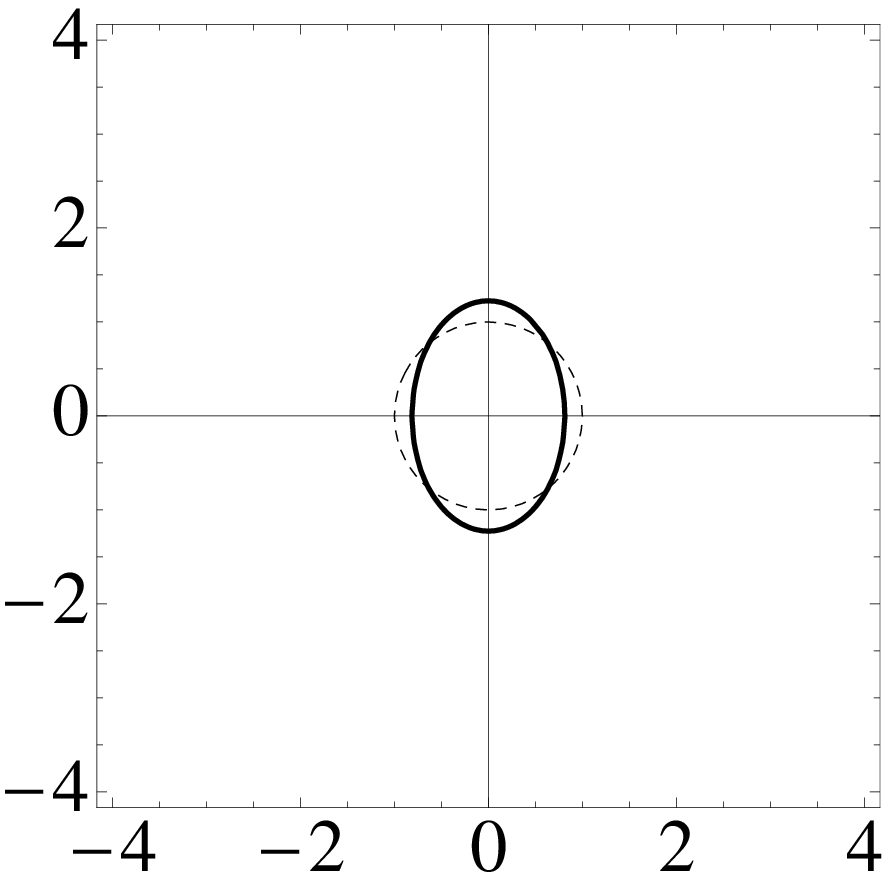}}
\subfloat[$t = 2 \pi$]{\label{fig:78}\includegraphics[width=0.313\textwidth]{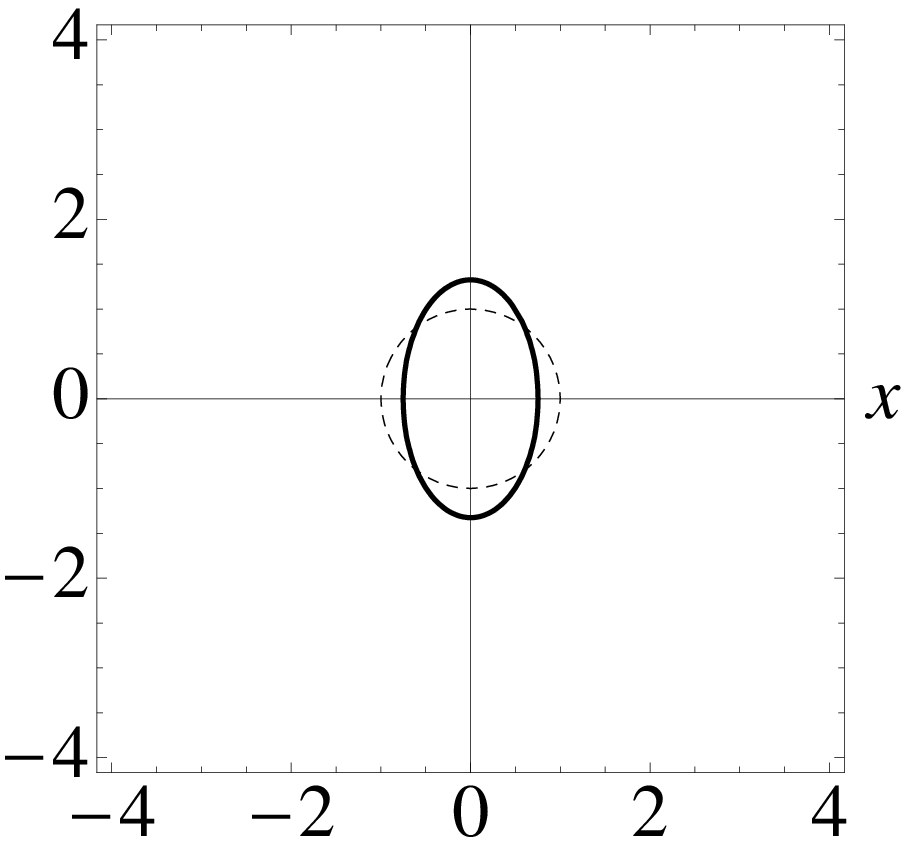}}
\caption{Time-evolution of the $\frac{1}{\rme}$-contour of  $W^{\textnormal {IP}}_{2}(x,p;t)$  given by (\ref{Wqu:Int}) in a time period of $2 \pi$ ($\sigma_x = 1, \sigma_p = \half, \om_0 =1, \kappa= 0.1$). The amount of the squeezing changes in time. We see that in the course of time the initially squeezed in the $p$-direction distribution function becomes squeezed in the $x$-direction due to the presence of the quadratic interaction term. The dashed circle represents the vacuum coherent  state.}
\label{fig.6}
\end{figure}

So far we have considered an interaction Hamiltonian that is linear in the creation and annihilation operators. From figures \ref{fig.1}--\ref{fig.5} it is apparent that such a Hamiltonian preserves the covariance matrix of the Gaussian distribution of the Wigner function, i.e., an initial coherent/ squeezed state remains coherent/squeezed under time evolution, although it might perform different types of rotations. Mathematically, this means the covariance matrix given by (\ref{covmatrix_int}) is invariant under a linear time-dependent interaction term. In this section we consider a quadratic interaction term in the Hamiltonian to illustrate the deformation of the covariance matrix. We show that in this case the contour of the Wigner function initially squeezed in the $p$-direction expands into a coherent state and then becomes squeezed in the $x$-direction. We choose the quadratic Hamiltonian to be
\BEA
\label{Hqu}
\hH(t) = \om_0 \had \ha + \rmi \kappa \left[ e^{ 2 \rmi \om_0 t} \ha^2-  e^{- 2 \rmi \om_0 t} (\had)^2  \right],
\EEA
where $\kappa$ is a real coupling constant. A physical realization of the above Hamiltonian may be the interaction of a coherent light beam with a non-linear optical medium. In that case, $\kappa$ contains the non-linear susceptibility and the pump field amplitude. The group-theoretical approach to the time-evolution of the Wigner function with a quadratic potential has been immensely studied~\cite{Wod,Gerry,Dattoli2,Dattoli1,Han,Orlowski,Vourdas,Zoubi}.
In what follows we use the Lie algebra of $\mbox{SU(1,1)}$\cite{Wod,Gerry} realized in terms of $\ha$ and $\had$,
\BEA\label{K0pm:def}
\K_0 = \half( \had \ha + \half),\qquad \K_+ = \half (\had)^2,\qquad \K_- = \half \ha^2,
\EEA
such that 
\BEA\label{K0pm:comm}
\left[\K_0, \K_{\pm} \right] = \pm \K_{\pm},\qquad \left[\K_-,\K_+ \right] = 2 \K_0.
\EEA
Thus in terms of these operators we have a Hamiltonian with linear interaction term
\BEA
\label{Hqu:K}
\hH(t) = 2 \om_0 \K_0  + \rmi \kappa\left(e^{ 2 \rmi \om_0 t} \K_- - e^{- 2 \rmi \om_0 t} \K_+\right).
\EEA
It is straight forward to calculate the unitary time-evolution operator in the interaction picture using (\ref{UMagnus}). In this case the interaction picture potential $\vi$ becomes independent of time and as a result the Magnus series contains only the first term $\hA_1(t)$ given by (\ref{Magnus}). Thus $\ui$ reads
\BEA\label{Uqu:int}
\ui = \rme^{\kappa t (\K_- - \K_+)}.
\EEA
$\ui$ resembles a squeezing operator with a time-dependent parameter, as a generalization of the time-dependent displacement operator in (\ref{UI}) for the linear interaction term in terms of $\ha$ and $\had$ calculated in the previous section.

\begin{figure}
\subfloat[$\om_0 t = 0$]{\label{fig:80}\includegraphics[width=0.3\textwidth]{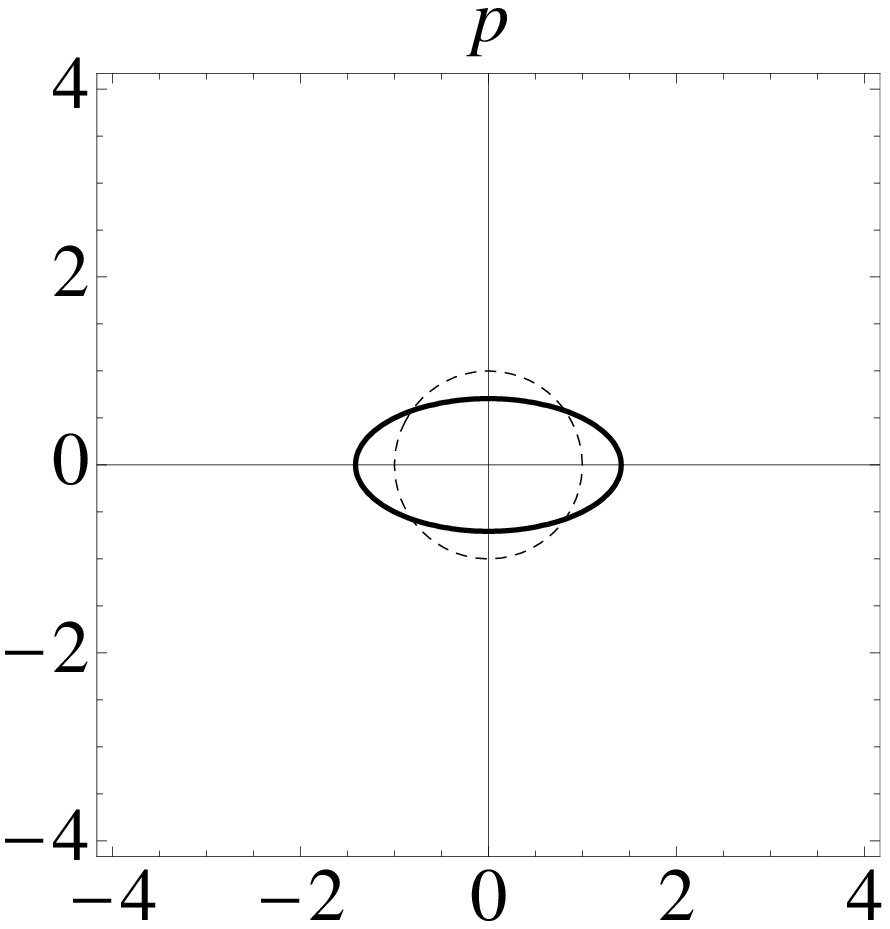}}
\subfloat[$\om_0 t = \frac{\pi}{4}$]{\label{fig:81}\includegraphics[width=0.3\textwidth]{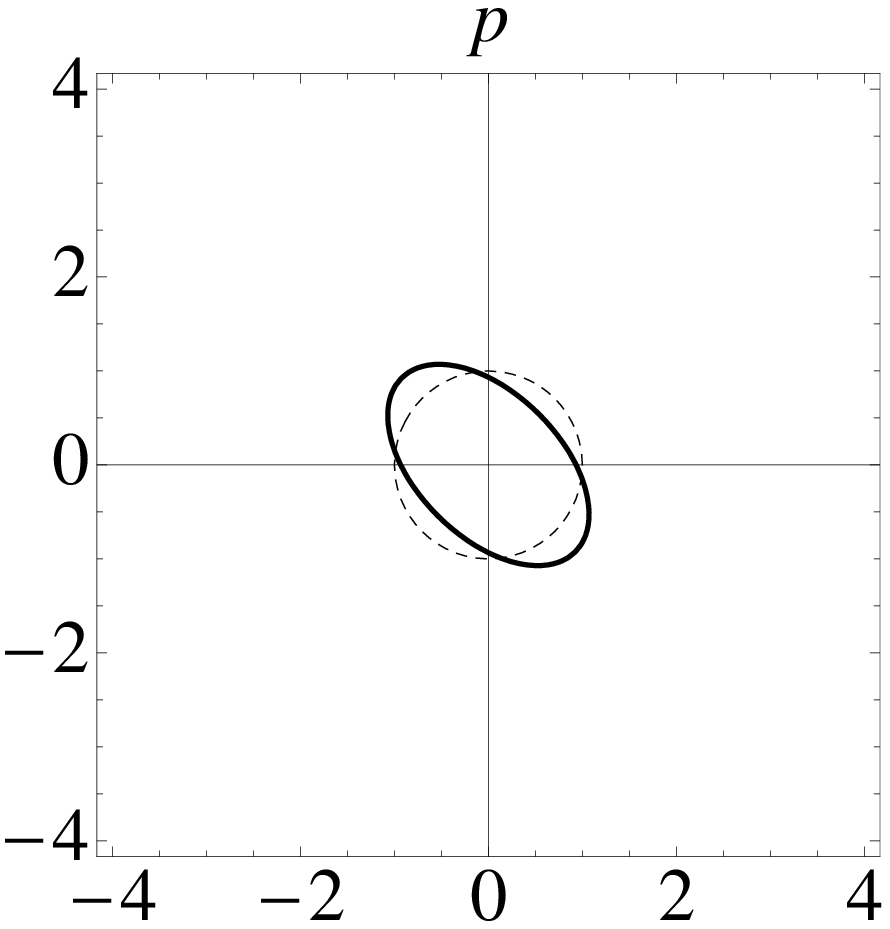}}
\subfloat[$\om_0 t = \frac{\pi}{2}$]{\label{fig:82}\includegraphics[width=0.313\textwidth]{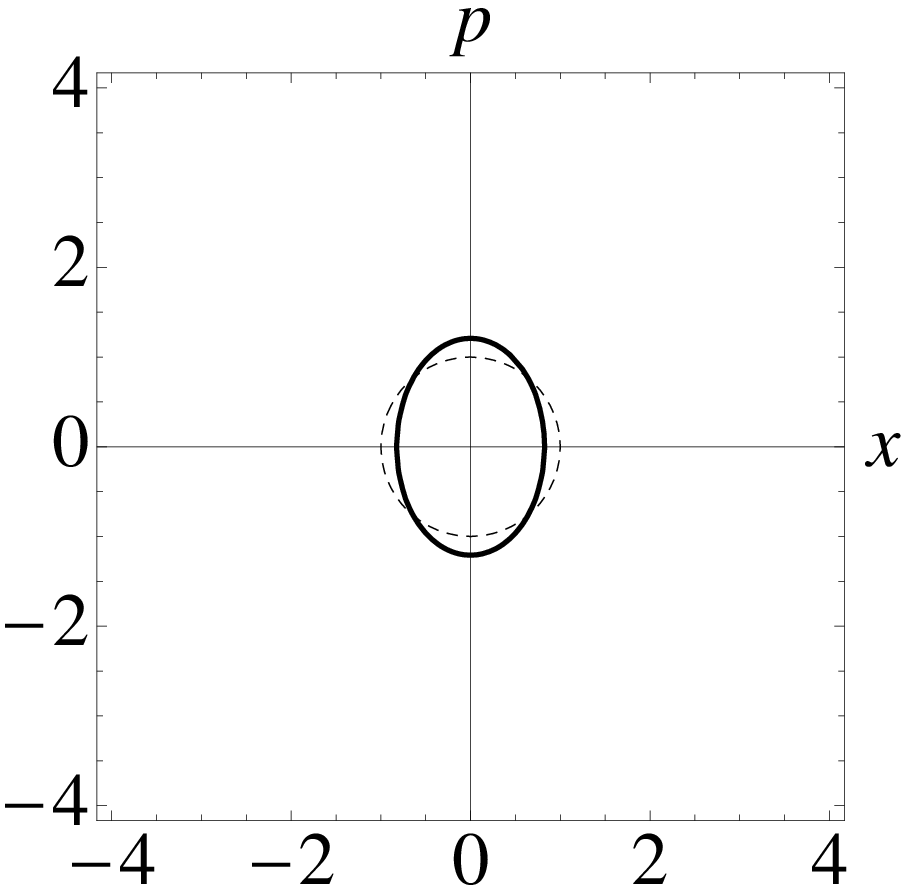}}\\
\subfloat[$\om_0 t= \frac{3 \pi}{4}$]{\label{fig:83}\includegraphics[width=0.3\textwidth]{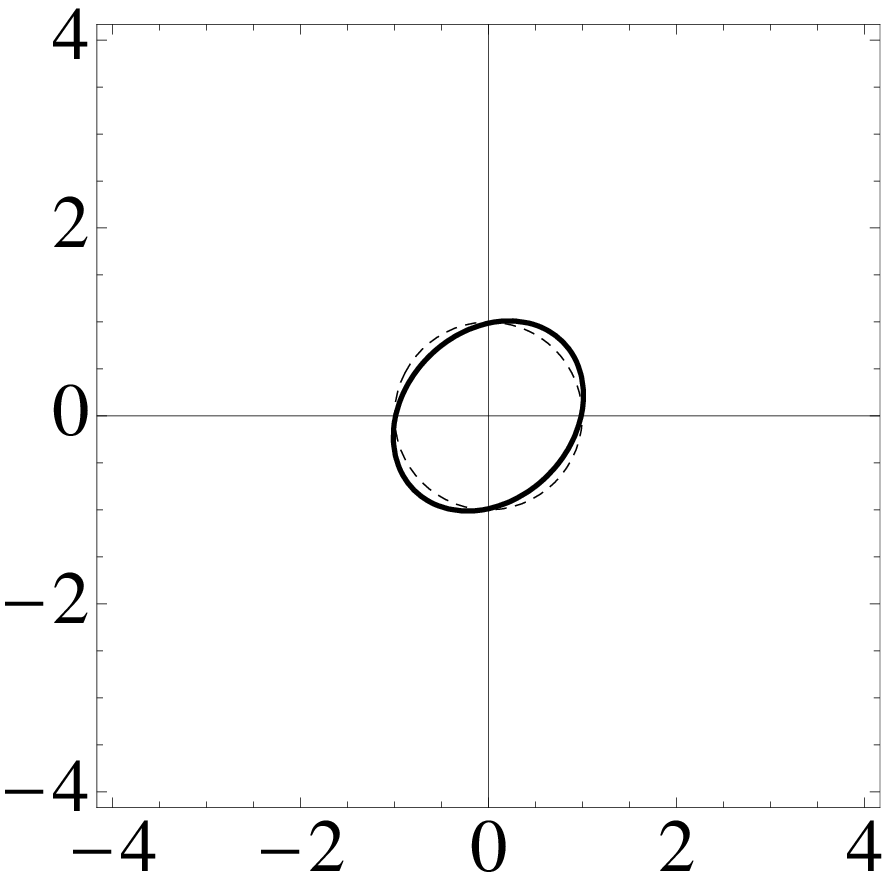}}
\subfloat[$\om_0 t = \pi$]{\label{fig:84}\includegraphics[width=0.3\textwidth]{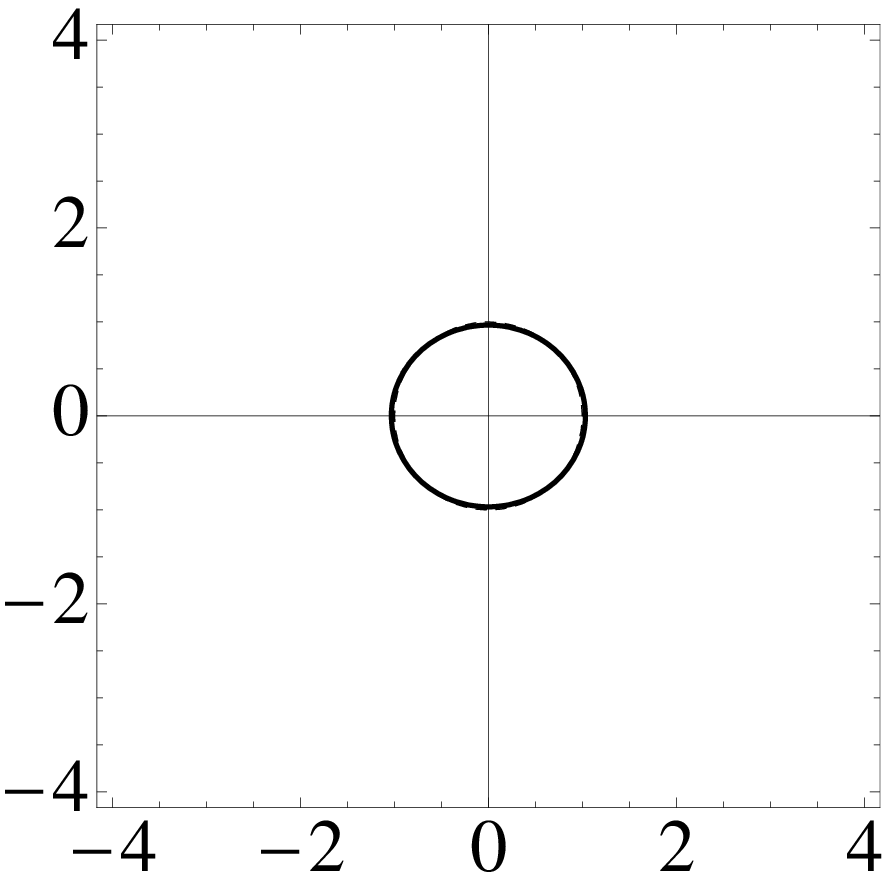}}
\subfloat[$\om_0 t = \frac{5 \pi}{4}$]{\label{fig:85}\includegraphics[width=0.313\textwidth]{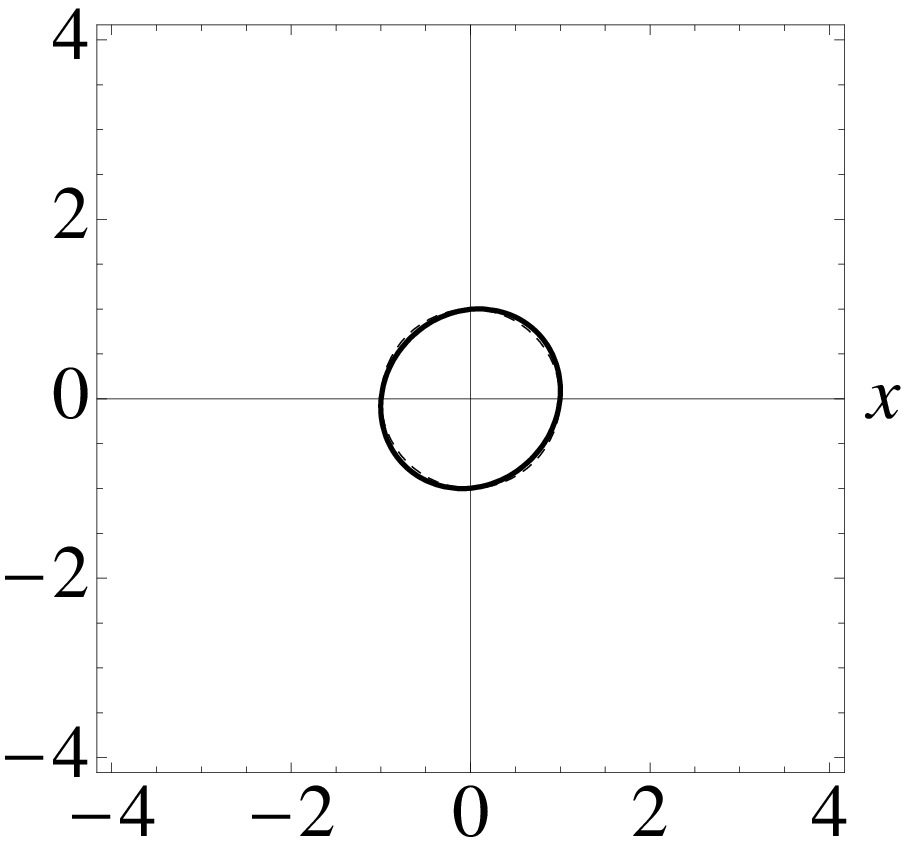}}\\
\subfloat[$\om_0 t = \frac{3 \pi}{2}$]{\label{fig:86}\includegraphics[width=0.3\textwidth]{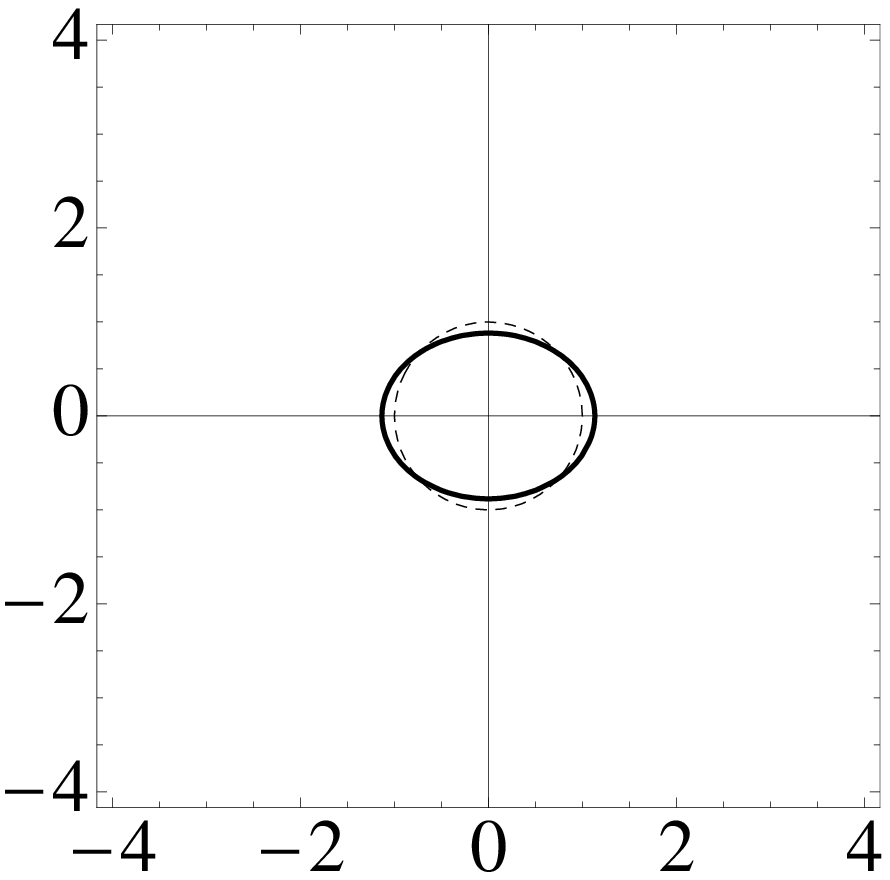}}
\subfloat[$\om_0 t = \frac{7 \pi}{4}$]{\label{fig:87}\includegraphics[width=0.3\textwidth]{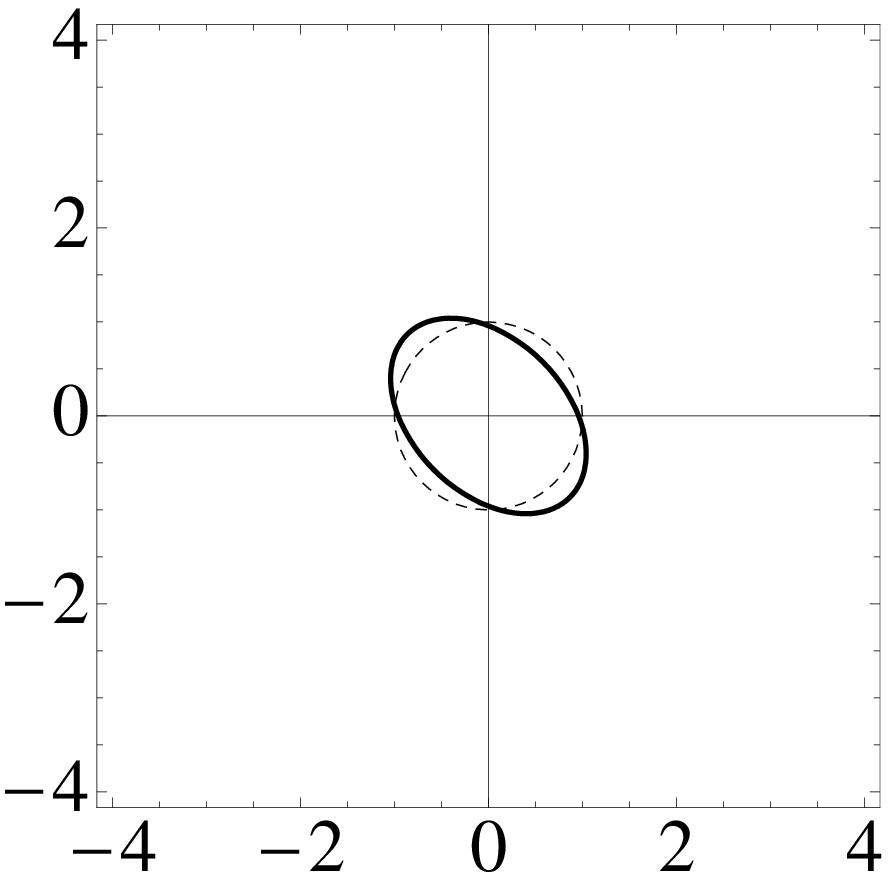}}
\subfloat[$\om_0 t = 2 \pi$]{\label{fig:88}\includegraphics[width=0.313\textwidth]{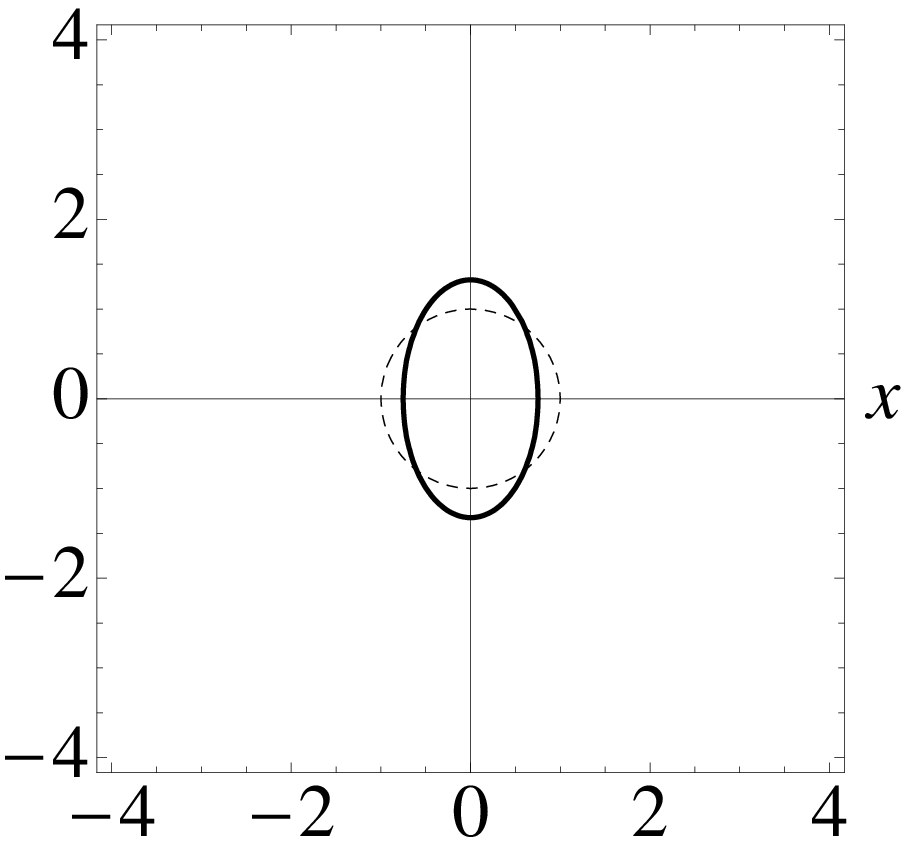}}
\caption{Time-evolution of the contour of $\frac{1}{\rme}$ value of $W^{\textnormal {SP}}_{2}(x,p;t)$  given by (\ref{Wqu:Sch}) in one period of $ \om_0 t$ ($\sigma_x = 1, \sigma_p = \half, \kappa= 0.1, \om_0 = 1$). The distribution function starts to expand in the $x$-direction and then squeeze in the $p$-direction. On top of that it also performs a clockwise rotation around the origin. The dashed circle represents the vacuum coherent state.}
\label{fig.7}
\end{figure}

Now we can calculate the state of the system at later time $t$ needed for the expectation values of the displacement operator. For the sake of simplicity, we assume the system starts its evolution in the IP from a vacuum squeezed state. This way, the centroid of the distribution coincides with the origin of the phase space. Then the symmetric characteristic function reads
\BEA
\label{chiqu:Int}
\Tr\left[\hroi (t) \D(\chi)\right] = \exp{\left\{- \frac{1}{8} \left[\frac{\xi^2}{\left(\sigma_p \rme^{\kappa t}\right)^2} + \frac{\eta^2}{\left(\sigma_x \rme^{- \kappa t}\right)^2}\right]\right\}}.
\EEA
We immediately realize that in the presence of a quadratic interaction the elements of the covariance matrix given by (\ref{covmatrix_int}) undergo a scaling (conformal) transformation. Furthermore, at a certain time the two elements of the covariance matrix become identical thus a squeezed states transforms into a coherent state.
Substituting (\ref{chiqu:Int}) into the definition of the Wigner function given by (\ref{Wignerdef}) and performing two Gaussian integrals over $\xi$ and $\eta$ gives the IP-Wigner function for the quadratic Hamiltonian as
\BEA
\label{Wqu:Int}
\Wip_2(x,p;t) = \frac{1}{\pi} \exp{\left[-\half \bi r^T\cdot \Xi ^{-1}_2(t)\cdot \bi r\right]},
\EEA
where the index $2$ represents the quadratic interaction term added to the free Hamiltonian. $\Xi^{- 1}_2(t)$ is the ``breathing'' covariance matrix defined as
\BEA\label{covmatrixqu:int}
\Xi^{-1}_2(t) = \left(\begin{array}{cc}\frac{\rme^{2 \kappa t}}{\sigma^2_x} & 0 \\0 &\frac{\rme^{-2 \kappa t}}{\sigma^2_p}\end{array}\right).
\EEA
This is illustrated in figure~\ref{fig.6}.

As it is showed in section~\ref{Sec:back} the transformation of the time-dependent density matrix back to the SP, amounts to a global clockwise rotation of the contour of the SP-Wigner function in phase space. In the case of quadratic interaction term this is straight forward to prove and yields to
\BEA\label{Wqu:Sch}
\Wsp_2(x,p;t) = \frac{1}{\pi} \exp{-\half \left[\bi r^T\cdot\Gamma^{-1}_2(t) \cdot \bi r\right]},
\EEA
where $\Gamma^{-1}_2(t)$ is defined as the clockwise rotated breathing covariance matrix,
\BEA\label{Gammaqu}
\Gamma^{-1}_2(t)\equiv R (- \om_0 t)\cdot \Xi ^{-1}_2(t) \cdot R(\om_0 t).
\EEA
This is shown in figure~\ref{fig.7}.
It is worth noticing that since the centroid of the Wigner function coincides with the origin of the phase space, the local counterclockwise rotation of the elliptical distribution around the centroid is canceled out by the global clockwise rotations around the origin as illustrated in figure \ref{fig.7}. However, this is not the case for an ideal squeezed state as it is illustrated in figure \ref{fig.1}.

\section{Summary}
We presented a pictorial representation of time evolution of a harmonic oscillator driven linearly and quadratically in \schro, \heis, and \inte\ picture (both \schro\ interaction picture, in which the evolution of the state is due to the interaction potential, and \heis\ interaction picture, where the the free evolution is contained in the state ). This has been done by employing the Wigner-Weyl representation to map the density matrices and observables in Hilbert space on distribution functions and variables, correspondingly, in phase space. As an example, we took the initial state of the system an ideal squeezed one. We showed that in the presence of a linear interaction term the time evolution of the corresponding Wigner function in the \schro\ picture amounts to different types of local and global rotations. The interaction picture time-evolution is mapped to a parallel transformation of the distribution function on a circle around the origin of the phase space. In transforming back to the \schro\ picture, we showed that the evolution of the Wigner function becomes more complicated due to the fact that it performs different types of rotations with difference frequencies. However, one can transform from one picture to another by performing passive and local transformations of the reference frame. Adding a quadratic interaction term to the Hamiltonian amounts to squeezing of the state with a time-dependent squeezing parameter. Classical picture of such interaction corresponds to the breathing of the Wigner function contours.


\section*{References}
\begin{thebibliography}{22}



\bibitem{Parisi} Auletta G and Parisi G 2001 {\it Foundations and Interpretation of Quantum Mechanics} (James Bennett Pty Ltd) p~ 48

\bibitem{Mandel} Mandel L and Wolf  E 1995 {\it Optical Coherence and Quantum Optics} (Cambridge University Press) 1042--45

\bibitem{Case08} Case W B 2008 ``Wigner functions and Weyl transforms for pedestrians'' {\it Am. J. Phys.} {\bf 76}, 937--46

\bibitem{Hillery} Hillery M, O'Connnell R F, Scully M O and Wigner E P  1984``Distribution functions in Physics: Fundamentals''  {\it Phys. Rep.}  {\bf 106}(3), 121--67

\bibitem{Agarwal} Agarwal G S 1987 ``Wigner-function description of quantum noise in interferometers'' {\it J. Mod. Opt.} {\bf 34}(6/7) 909--21

\bibitem{Schleich} Schleich W P 2001 {\it Quantum Optics in Phase Space} 1st. ed. (Wiley-VCH Verlag Berlin GmbH) 321--42

\bibitem{Kim90} Kim Y S and Wigner E P 1990 ``Canonical transformation in quantum mechanics'' {\it Am. J. Phys.} {\bf 58}  439--48


\bibitem{Kim88}Han D, Kim Y S and Noz M E 1988 ``Linear canonical transformations of coherent and squeezed states in the Wigner phase space'' {\it Phys. Rev.} A {\bf 37} 807--14

\bibitem{Kim89}Han D, Kim Y S and Noz M E 1989 ``Linear canonical transformations of coherent and squeezed states in the Wigner phase space. II. Quantitative analysis'' {\it Phys. Rev.} A {\bf 40}, 902--12


\bibitem{Glauber} Glauber R J 1963 ``Coherent and incoherent states of the radiation field'' {\it Phys. Rev.} {\bf 131} 2766--88

\bibitem{Barnett} Barnett S M and Radmore P M 1997 {\it Methods in Theoretical Quantum Optics} (Clarendon Press,Oxford) p~ 106

\bibitem{Scully} Scully M O and Zubairy M S 2002 {\it Quantum Optics} (Cambridge University Press) p~ 51

\bibitem{Yuen76}Yuen H P  1976 ``Two-photon coherent states of the radiation field'' {\it Phys. Rev.} A {\bf 13} 2226--43

\bibitem{Caves81} Caves C M 1981 ``Quantum-mechanical noise in an interferometer'' {\it Phys. Rev.} D {\bf 23} 1693--708


\bibitem{Walls83} Walls D F 1983  ``Squeezed states of light'' {\it Nature} {\bf 306} 141--6


\bibitem{Tapia93} Mu\~{n}oz-Tapia R  1993``Quantum mechanical squeezed state'' {\it Am. J. Phys.} {\bf 61} 1005--8

\bibitem{Magnus} Magnus W 1954 ``On the exponential solution of differential equations for a linear operator''  {\it Communications on Pure and Applied Mathematics} {\bf 7} 649--73. For a general review on the Magnus series both from the Mathematical and the Physical point of view see Blanes S, Casas F, Oteo J A and Ros J  2009 ``The Magnus expansion and some of its applications'' {\it Phys. Rep} {\bf 470} 151--238



\bibitem{Roulette} Walker G 1939 The theory of roulettes and glissettes
{\it National Mathematics Magazine} {\bf 13} No. 5, 223--29

\bibitem{SHIP} Barnett S M and Radmore P M 1997 {\it Methods in Theoretical Quantum Optics} (Clarendon Press,Oxford) p~ 14

\bibitem{Wod} W\`{o}dkiewicz K and Eberly J H  1985 ``Coherent states, squeezed fluctuations, and the SU(2) and SU(1,1) groups in quantum-optics applications'' {\it J. Opt. Soc. Am. B} Mar 1;{\bf2(3)} 458--66. 

\bibitem{Gerry} Gerry C C 1985 ``Dynamics of SU(1,1) coherent states'' {\it Phys. Rev. A} {\bf31} 2721--23.

\bibitem{Dattoli2} Dattoli G and Torre A. 1988 ``Algebraic view to the quantum anharmonic oscillator'' {\it Phys. Rev. A} {\bf 37(5)} pp.1571--75.

\bibitem{Dattoli1} Dattoli G, Solimeno S, Torre A. 1987  `` Algebraic view of the optical propagation in a nonhomogeneous medium''. {\it Phys. Rev. A.} {\bf 35} 1668--72. 


\bibitem{Han} Han D, Kim YS, Noz ME. 1989 ``Linear canonical transformations of coherent and squeezed states in the Wigner phase space. II. Quantitative analysis''  {\it Phys. Rev. A.} {\bf 40(2)} 902--12. 

\bibitem{Orlowski} Orlowski A, W\'{o}dkiewicz K  1990 ``On the SU(1, 1) Phase-space Description of Reduced and Squeezed Quantum Fluctuations''  {\it Journal of Modern Optics}  {\bf 37(3)} 295--301. 

\bibitem{Vourdas} Vourdas A 1990  ``SU(2) and SU(1,1) phase states'' {\it Phys. Rev. A.}  {\bf 41}1653--61. 



\bibitem{Zoubi} Zoubi H and Ben-Aryeh Y. 1998 ``The evolution of harmonic oscillator Wigner functions described by the use of group representation'' {\it Quantum Semiclass. Opt.} {\bf 10} 447--58.

\end{thebibliography}
\end{document}